%% file: main.tex
\documentclass[journal]{vgtc}                 

\ifpdf
  \pdfoutput=1\relax                   
  \usepackage{graphicx}                
  \DeclareGraphicsExtensions{.pdf,.png,.jpg,.jpeg} 
\else
  \usepackage{graphicx}                
  \DeclareGraphicsExtensions{.eps}     
\fi%

\graphicspath{{figures/}{pictures/}{images/}{./}} 

\usepackage{microtype}                 
\PassOptionsToPackage{warn}{textcomp}  
\usepackage{textcomp}                  
\usepackage{mathptmx}                  
\usepackage{times}                     
\usepackage{cite}                      
\usepackage{tabu}                      
\usepackage{booktabs}                  

\usepackage{mathptmx}
\usepackage{graphicx}
\usepackage{comment}
\usepackage{subfigure} 

\usepackage{algpseudocode}
\usepackage{algorithm}
\usepackage{enumitem}

\usepackage{subfiles}
\usepackage{graphicx}
\usepackage{subfigure}
\usepackage{epstopdf}
\usepackage{float}

\usepackage{etoolbox}
\usepackage[mathscr]{euscript}
\usepackage{etoolbox}
\usepackage{setspace}
\usepackage{dingbat}
\usepackage{xspace}

\usepackage{amsmath,amsthm,amssymb}
\usepackage[table]{xcolor}
\usepackage{moresize}
\usepackage[utf8]{inputenc}
\usepackage{wrapfig}
\usepackage{multirow}
\usepackage{rotating}
\usepackage[normalem]{ulem}
\usepackage{tabularx}

\newcommand {\mm}[1] {\ifmmode{#1}\else{\mbox{\(#1\)}}\fi}

\newcommand{\Fgroup}        {{\mathcal{F}}}

\newcommand{\lnorm}        {\ell}

\newcommand{\BigO} {\mathcal{O}}

\usepackage{tikz}

\newcommand{\methodTDA}{\textsc{topology}\xspace}
\newcommand{\methodUni}{\textsc{uniform}\xspace}
\newcommand{\methodGau}{\textsc{Gaussian}\xspace}
\newcommand{\methodMin}{\textsc{min}\xspace}
\newcommand{\methodMax}{\textsc{max}\xspace}
\newcommand{\methodMed}{\textsc{median}\xspace}
\newcommand{\methodMean}{\textsc{mean}\xspace}
\newcommand{\methodCut}{\textsc{cutoff}\xspace}
\newcommand{\methodDP}{\textsc{Douglas-Peucker}\xspace}
\newcommand{\methodSG}{\textsc{Savitzky-Golay}\xspace}
\newcommand{\methodButt}{\textsc{Butterworth}\xspace}
\newcommand{\methodCheb}{\textsc{Chebyshev}\xspace}

\onlineid{1250}

\vgtccategory{Research}
\vgtcpapertype{Empirical Study}

\vgtcinsertpkg

\title{LineSmooth: An Analytical Framework for Evaluating the Effectiveness of Smoothing Techniques on Line Charts}

\shortauthortitle{Rosen and Quadri: A Framework for Evaluating Line Chart Smoothing}

\author{Paul Rosen and Ghulam Jilani Quadri}
\authorfooter{
\item
 Paul Rosen is with the University of South Florida. E-mail: prosen@usf.edu.
\item
 Ghulam Jilani Quadri is with the University of South Florida. E-mail: ghulamjilani@usf.edu.
}

\teaser{
    \centering
    \includegraphics[width=0.935\linewidth]{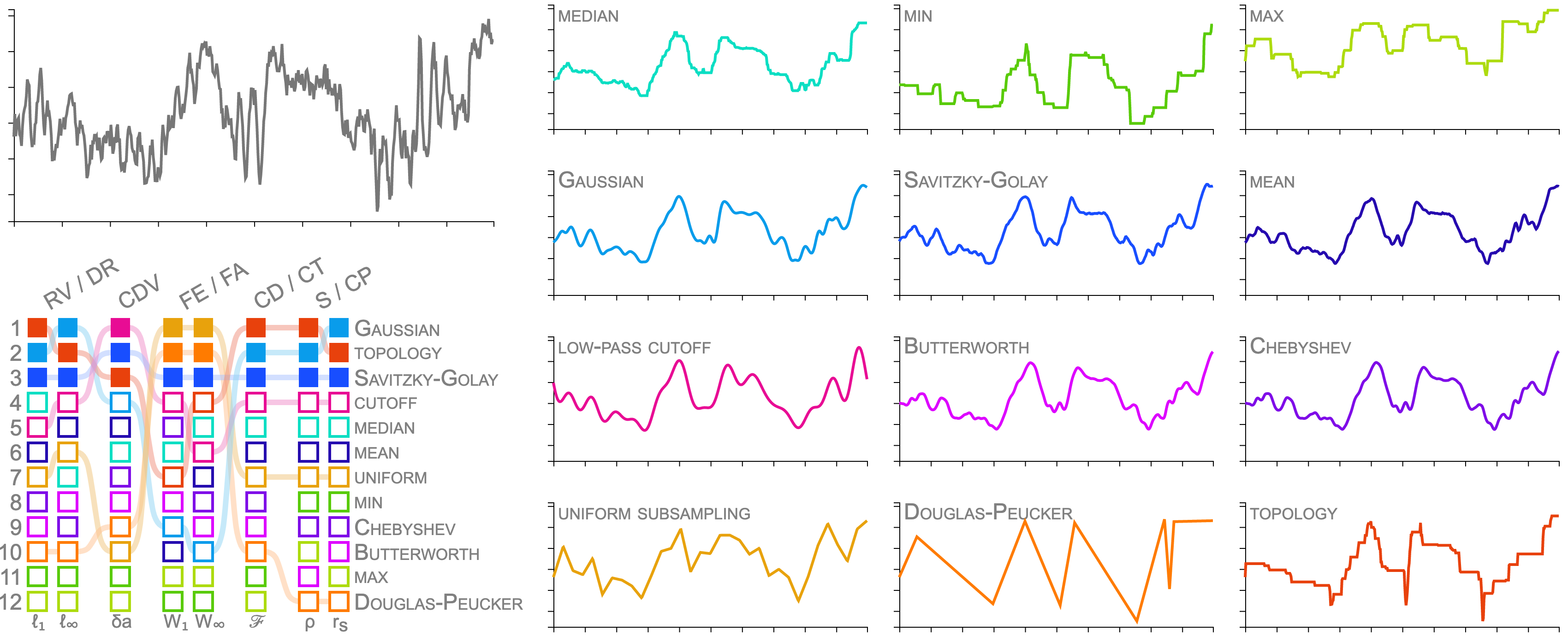}
    \begin{minipage}[t]{0pt}
        \hspace{-430pt}
        \begin{minipage}[t]{0.6\textwidth}
            \vspace{-120pt}
            \subfigure[Input Data\label{fig:teaser:input}]{\hspace{125pt}}
            
            \vspace{89pt}
            \subfigure[Effectiveness Rank By Task\label{fig:teaser:ranks}]{\hspace{108pt}}
        \end{minipage}
    \end{minipage}    
    \begin{minipage}[t]{0pt}
        \hspace{-204pt}
        \begin{minipage}[t]{0.6\textwidth}
            \vspace{-145pt}
            \subfigure[Rank-based Smoothing\label{fig:teaser:rank}]{\hspace{120pt}}
            
            \vspace{22pt}
            \subfigure[Convolutional Smoothing\label{fig:teaser:conv}]{\hspace{120pt}}
            
            \vspace{21pt}
            \subfigure[Frequency Domain Smoothing\label{fig:teaser:freq}]{\hspace{120pt}}
            
            \vspace{21pt}
            \subfigure[Subsampling\label{fig:teaser:subs}]{\hspace{120pt}}
        \end{minipage}
    \end{minipage}

    \caption{Example of the (a) EEG Channel 10 (500 samples) dataset with (c-f) 4 classes / 12 techniques of line chart smoothing applied. (b)~Our framework provides the capability to rank smoothing methods by their efficacy in visual analytics tasks. To form the rankings, the 12 smoothing examples are calibrated to have similar visual complexity. Then, 8 measures of effectiveness, described in \autoref{sec:measures}, are calculated and ordered from best to worst. The results show that for this example, the \methodGau, \methodTDA, and \methodSG methods are the 3 best techniques for the Retrieve Value (RV), Determine Range (DR), Characterize Distribution (CD), Cluster Trends (CT), Sort (S), and Cluster Points (CP) tasks. For the Compute Derived Value (CDV) task, \methodCut, \methodSG, and \methodTDA perform best, respectively. Finally, for the Find Extrema (FE) and Find Anamolies (FA) tasks, \methodUni subsampling, \methodDP, and \methodSG perform best. Clearly, no single method is best for all visual analytics tasks, but for this particular data, \methodSG, being in the top 3 for all tasks, would be a reasoned choice.}
    \label{fig:teaser}
}

\input{sec-abstract}

\begin{document}


\firstsection{Introduction}

\maketitle


\input{sec-intro}

\input{sec-related}

\input{sec-taxonomy}

\input{sec-framework-metrics}

\input{sec-framework-tasks}

\input{sec-framework-eval}

\input{sec-results}

\input{sec-discussion}

\acknowledgments{We would like to thank Bei Wang and Ashley Suh for providing valuable input on this project. This work was supported in part by a grant from the National Science Foundation (IIS-1845204).}

\bibliographystyle{abbrv-doi}
\bibliography{main}

\end{document}

%% file: sec-abstract.tex
\abstract{We present a comprehensive framework for evaluating line chart smoothing methods under a variety of visual analytics tasks. Line charts are commonly used to visualize a series of data samples. When the number of samples is large, or the data are noisy, smoothing can be applied to make the signal more apparent. However, there are a wide variety of smoothing techniques available, and the effectiveness of each depends upon both nature of the data and the visual analytics task at hand. To date, the visualization community lacks a summary work for analyzing and classifying the various smoothing methods available. In this paper, we establish a framework, based on 8 measures of the line smoothing effectiveness tied to 8 low-level visual analytics tasks. We then analyze 12 methods coming from 4 commonly used classes of line chart smoothing---rank filters, convolutional filters, frequency domain filters, and subsampling. The results show that while no method is ideal for all situations, certain methods, such as \methodGau filters and \methodTDA-based subsampling, perform well in general. Other methods, such as low-pass \methodCut filters and \methodDP subsampling, perform well for specific visual analytics tasks. Almost as importantly, our framework demonstrates that several methods, including the commonly used \methodUni subsampling, produce low-quality results, and should, therefore, be avoided, if possible.}

\keywords{Line chart, data smoothing, time-series.}

%% file: sec-intro.tex
Line charts, which date back to William Playfair~\cite{playfair1801commercial}, are commonly used for visualizing time-series and continuous data. Borkin et al.\ found that line charts are the second most frequently used visualization type, only behind bar charts, in scientific publications, news media, government, and world organizations materials~\cite{borkin2013makes}.

When line chart data are noisy, e.g., see~\autoref{fig:teaser:input}, visualization designers can turn to \textit{smoothing} to reduce the visual clutter. However, there are many techniques available (see \autoref{fig:teaser}(c-f)), and while the results they produce may look similar, each preserves different properties of the data. For example, rank-based (see \autoref{fig:teaser:rank}) and convolutional smoothing methods (see \autoref{fig:teaser:conv}) preserve local properties, such as local trends, while frequency-domain smoothing (see \autoref{fig:teaser:freq}) and subsampling (see \autoref{fig:teaser:subs}) preserve global properties, such as the most prominent peaks in the data.

To preserve some properties of the input data, each smoothing technique must also \textit{lose information}, which can have a negative impact on the utility of the resulting data. To further complicate matters, the importance of the lost information can be influenced by both the data being used and the visual analytics tasks being performed. To date, the visualization community lacks a comprehensive framework for measuring the influence of line chart smoothing on a range of visual analytics tasks.

In this paper, we present an analytical framework for measuring the effectiveness of various smoothing techniques under 8 different low-level visual analytics tasks performed on line charts. We define a taxonomy of 4 classes of line chart smoothing and evaluate a total of 12 commonly available techniques on 80 datasets from 13 categories.

Through our analysis, we show that there is no single smoothing technique that is ideal for all visual analytics tasks. Furthermore, we show that the efficacy of each technique can vary by the datasets being analyzed. Nevertheless, we identify specific methods that consistently perform well, in particular \methodGau filters and \methodTDA-based subsampling~\cite{suh2019topolines}. In other cases, some methods are particularly well suited for specific tasks, e.g., low-pass \methodCut filters and \methodDP subsampling~\cite{douglas1973algorithms} are well suited for computing a derived value and finding extrema, respectively. Finally, we identify several methods, including the commonly used \methodUni subsampling, which perform consistently poorly.

Visualization designers can use this framework and results of this paper to either: (1)~select a smoothing technique, which is most effective in general or most effective for the tasks their users perform; (2)~evaluate their data to select the technique that is specifically most effective; or (3)~to understand how much error is introduced as they increase the level of smoothing used in their visualizations. An interactive version of our framework is at {\small\textless\textcolor{blue}{\url{https://usfdatavisualization.github.io/LineSmoothDemo}}\textgreater}.

%% file: sec-related.tex
\section{Prior Work}

We discuss prior work in the context of analytical tasks performed using line charts, decision-making with line charts, and line chart smoothing.

\subsection{Task Efficacy}

Line charts, which are traditionally used to visualize time-series and
continuous 1D data~\cite{1986tufte,1989kosslyn,Zacks1999BarsAL}, have been studied in the context of a variety of low-level visual analytics tasks~\cite{amar2005low}. A recent multi-chart experimental study found that line charts are significantly more accurate than other charts for the tasks of correlation and, to a lesser extent, finding extrema, characterizing distributions, and filtering~\cite{2018saket}. Even so, line charts are used for a wider variety of visual analytics tasks.

\paragraph{Comparison Tasks}
%
Early work on horizon graphs~\cite{2005horizon} investigated their effectiveness as compared to line charts in a comparison task~\cite{2009heer}. The work identified space-accuracy trade-off that could be used to optimize perception between the two.
%
Another study compared line charts to horizon graphs and colorfields for similarity assessment~\cite{2019gogolou}. The study showed that deformations in the data are perceived differently depending on the visualization, and, in particular, line charts are more sensitive to changes in amplitude than position.

\paragraph{Statistical Tasks}
%
Line charts are used in many forms of statistical analysis~\cite{nourbakhsh1994statistical}.
%
Perception-based experiments that measure user's judgments concluded that line charts have low-to-medium precision on estimating correlation~\cite{kay2016beyond, harrison2014ranking}.
%
More generally, when considering aggregation tasks, it has been shown that line charts are effective at finding minima and maxima and determining value range while falling short on determining the average, spread, and outliers in the data~\cite{albers2014task}.
%
More specifically, when calculating averages, colorfields have been shown to outperform line charts~\cite{correll2012comparing}.

\paragraph{Trend Assessment}
Another common task attended to with line charts is trend assessment.
%
It has been shown that line charts are, in general, better at trend assessment than scatterplots and bar charts, particularly for nonlinear trends~\cite{best2007perception}. 
%
However, when outliers are introduced into the data, the trends in their estimates begin to diverge from standard regression models~\cite{correll2017regression}. 
%
Furthermore, when data are noisy, trends in the data are easier to identify using scatterplots~\cite{wang2018line}.

\paragraph{Visual Encoding, Layout, and Interaction}
The visual encodings, layout, and interaction with line charts can have an impact on their efficacy. 
%
For example, color is an important visual encoding. For line charts, it has been shown that color difference varies inversely with thickness~\cite{szafir2018modeling}. In other words, to be effective, a light-colored line must be thicker than dark-colored ones. 
%
Concerning layout, the efficacy of line charts is subject to the choice of aspect ratio, which can be automatically optimized for a chart~\cite{2018aspectratio}. 
%
Javed et al.\ evaluated the effectiveness of line charts with small multiples, horizon graphs, stacked graphs, and braided graphs for comparison, slope, and discrimination tasks~\cite{2010javed}. The results showed that techniques with separate charts performed better for data with large visual spans, while shared-space techniques were better for short spans. 
%
A more recent study showed that overlaid line charts perform better than small multiples in comparison tasks~\cite{ondov2018face}.
%
Finally, adding user interaction to line charts can enhance the user experience without a loss to efficacy~\cite{2016adnan}.

\subsection{Decision-Making} 

Line charts have been studied in several decision-making scenarios as well.
%
In time-sensitive application settings, the ability to accurately interpret a line chart ``at a glance'' is crucial. Recently, Pixel Approximate Entropy (PAE) was used as a metric for the perceptual complexity of line charts, and it was shown that increased chart PAE correlates with reduced judgment accuracy~\cite{2018entropy}.

\paragraph{Missing Data} 
%
Another decision-making challenge in visualization is when data are missing; one needs to impute the missing data into the visualization~\cite{2012templ}.
%
An early study that looked at the problem of missing data in line charts on trend and comparison tasks found that even with missing data, user performance was high~\cite{2005eaton}.
%
One way to address missing data is to use additional visual channels, e.g., color, empty points, or error bars, which have been shown to improve analysis performance and confidence of users on average and trend finding tasks~\cite{2018szafir}.

\paragraph{Distortion} 
Another concern for decision-making is distortion in the visualization, such as an inverted axis or a distorted aspect ratio. Such situations can demonstrate a reversal of messaging, which can lead viewers to draw false inferences and judgments~\cite{pandey2015deceptive}.
%
Another example is when missing data are misleadingly inserted into a visualization, e.g., assigned arbitrary values, user performance can go down significantly~\cite{2005eaton}.
%
To address this weakness, multi-view systems have been proposed to assist in time-series data quality checking~\cite{2016arbesser}.

\subsection{Line Chart Smoothing}  
Smoothing line charts can be considered a form of distortion, as the data are being distorted to improve clarity. There has been prior work looking at smoothing in the signal processing community, e.g., Shao et al.\ compared 5 smoothing methods for vegetation classification~\cite{shao2016evaluation}, and image processing community, e.g., Chen and Yeh developed a quantitative evaluation for edge-preservation in image smoothing~\cite{chin1983quantitative}. To our surprise, we were unable to find any prior studies that evaluated the impact of various smoothing techniques to line charts, except for our own small-scale study that introduced a topology-based smoothing method~\cite{suh2019topolines}. Nevertheless, no comprehensive framework and evaluation, such as the one we are introducing in this paper, exists.

%% file: sec-taxonomy.tex
\begin{figure*}[!t]
    \centering

    \centering
    \subfigure[Rank Filter: \methodMed\label{fig:filter_illustrations:median}]{\includegraphics[height=78pt]{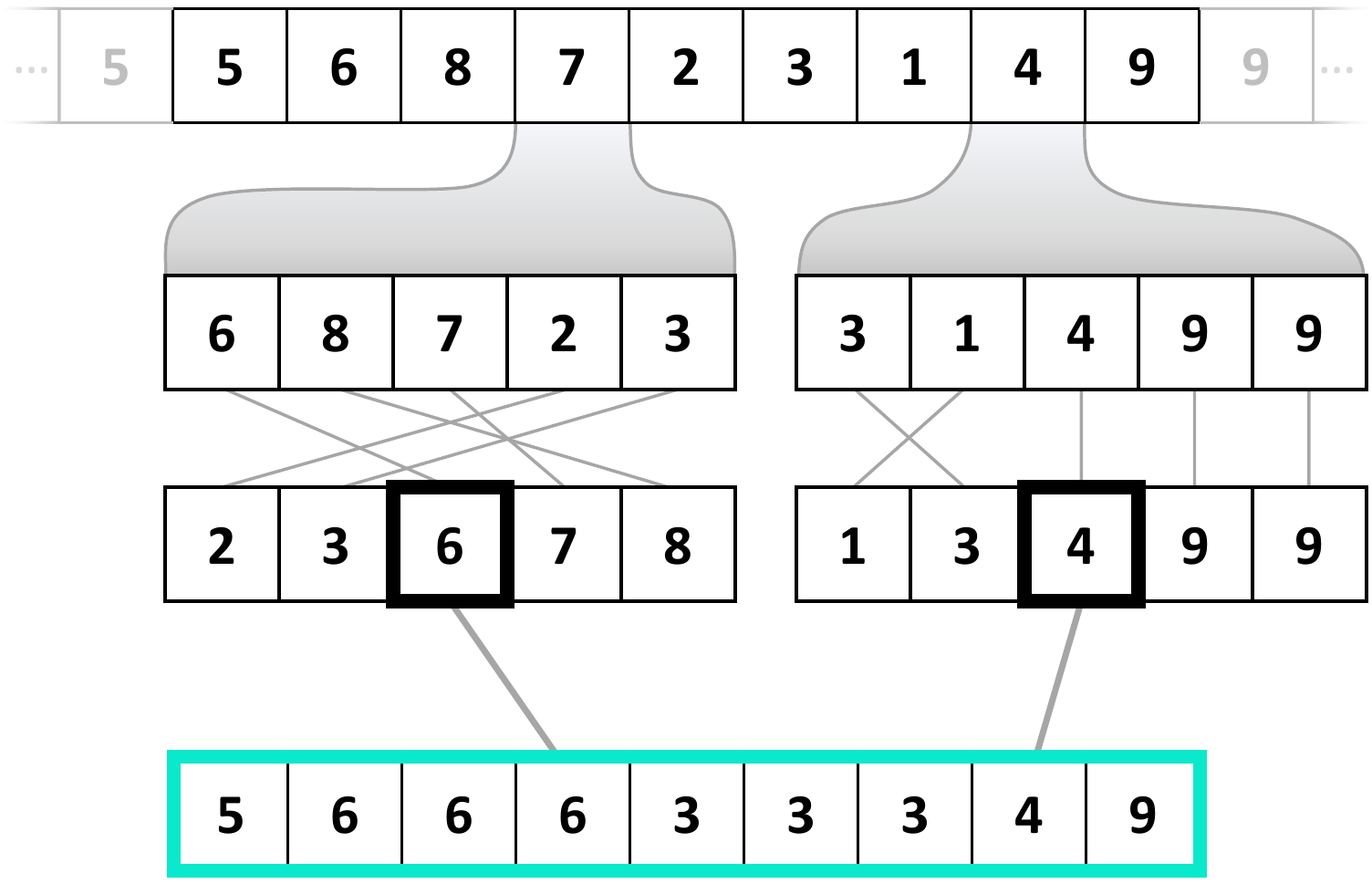}}
    \subfigure[Convolutional Filter: \methodGau\label{fig:filter_illustrations:gaussian}]{\includegraphics[height=78pt]{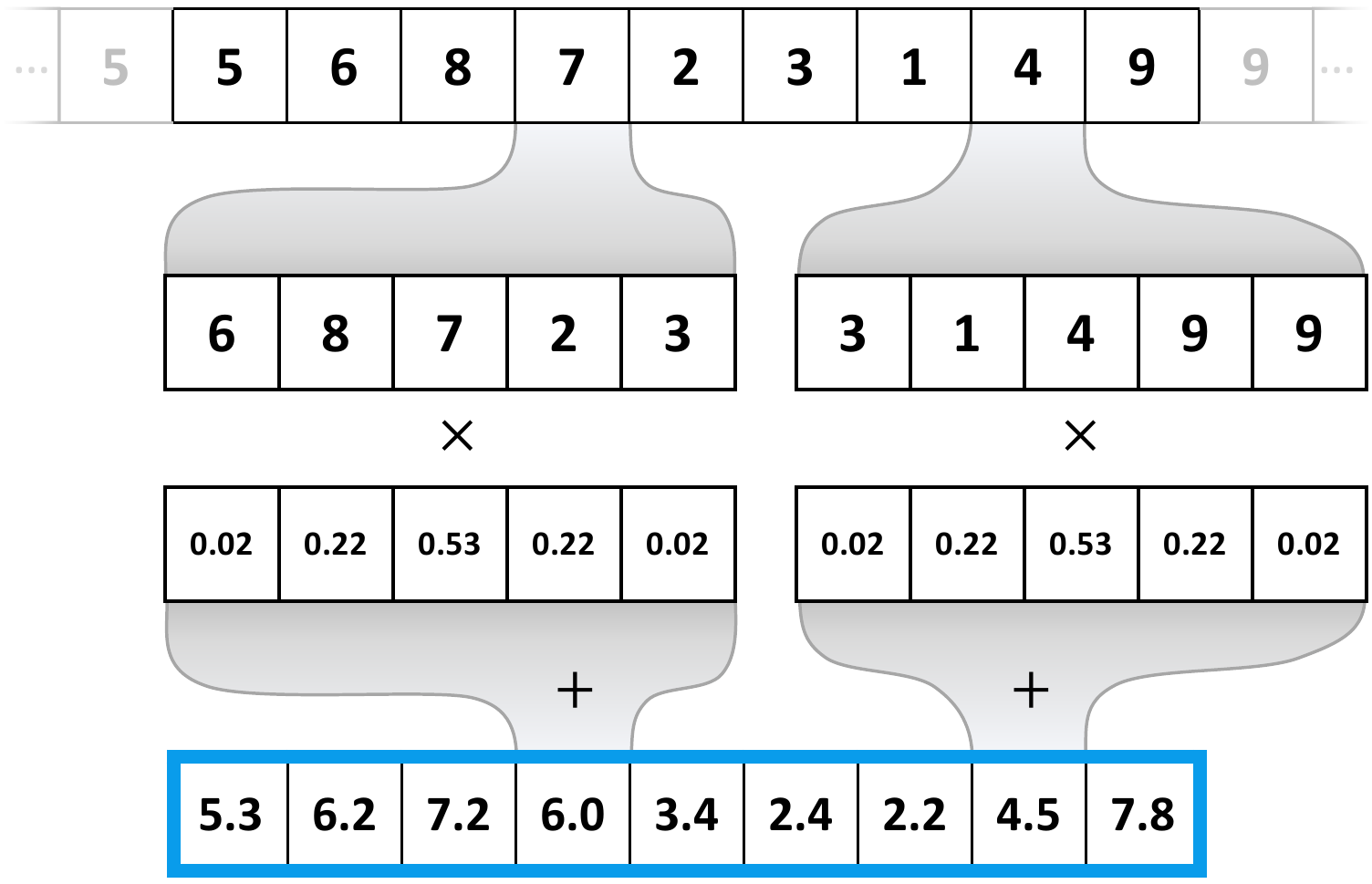}}
    \hfill
    \subfigure[Frequency Domain: \methodCut filter\label{fig:filter_illustrations:fft}]{\hspace{18pt}\includegraphics[height=78pt]{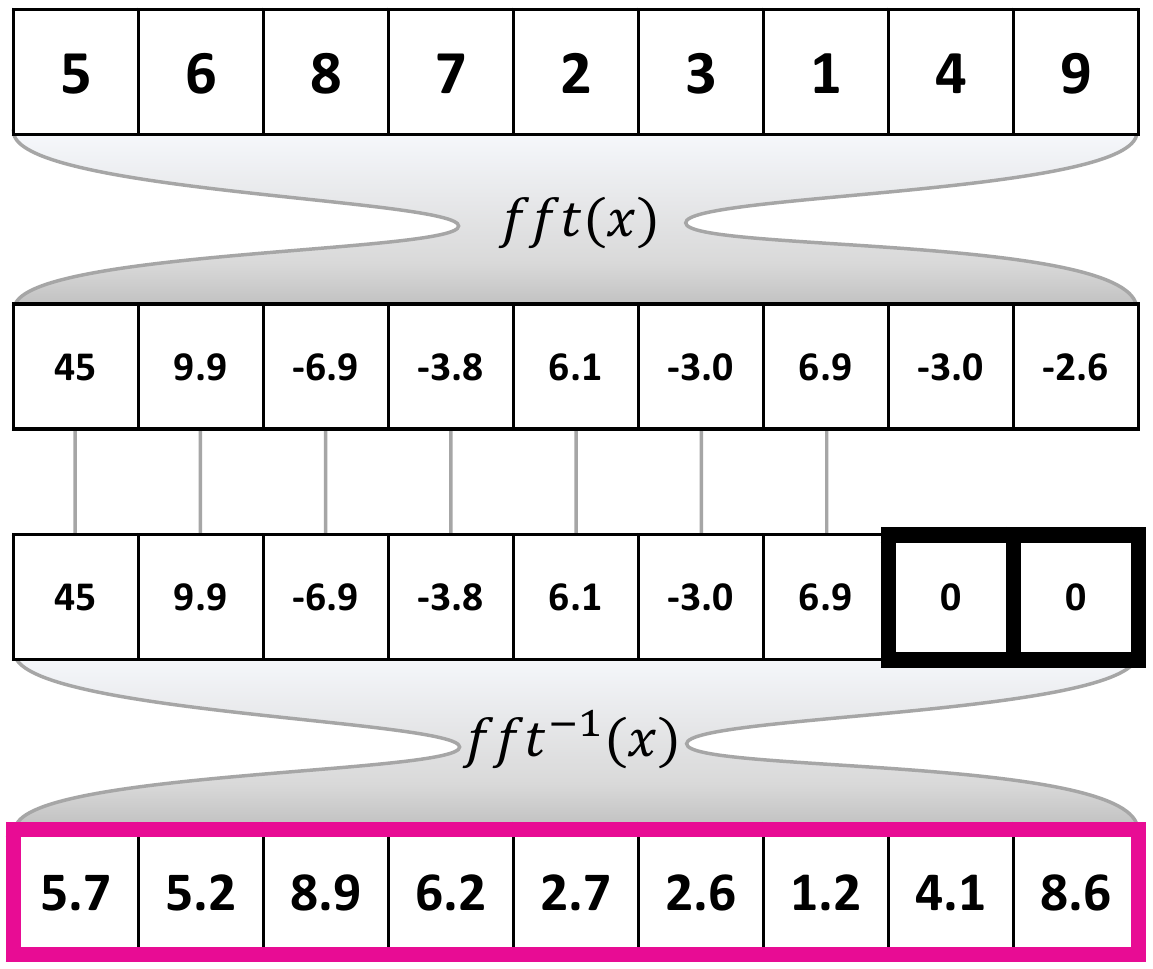}\hspace{18pt}}
    \hfill
    \subfigure[Subsampling: \methodDP\label{fig:filter_illustrations:dpr}]{\hspace{17pt}\includegraphics[height=78pt]{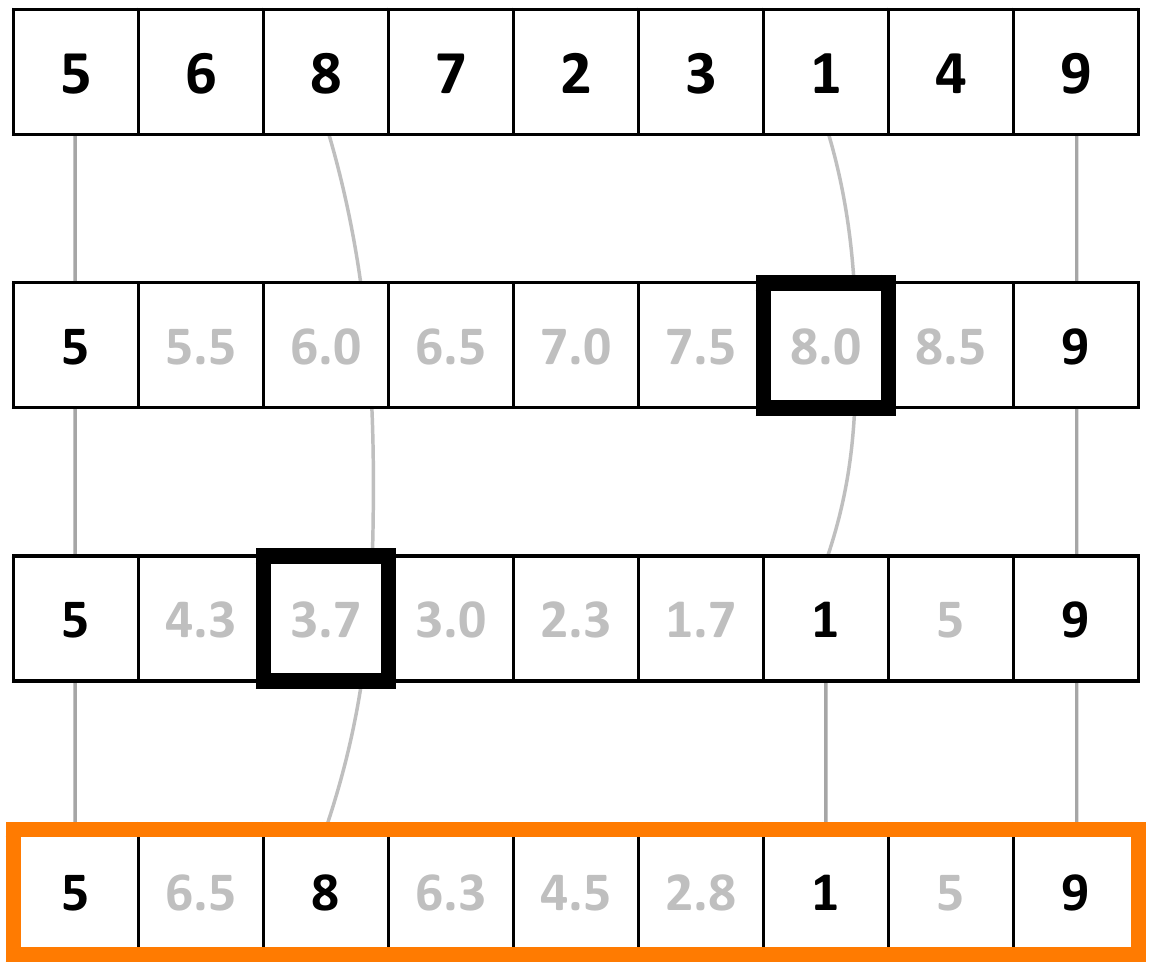}\hspace{17pt}}

    \caption{Illustration of (a-b) local and (c-d) global smoothing methods for line charts. 
    (a)~Starting with the input data (top), the \methodMed filter extracts a window (2nd row), sorts the window (3rd row), and selects the median value for output (4th row). 
    (b)~The \methodGau filter similarly extracts a local window (2nd row). However, the window has a convolution applied based upon a normal distribution (3rd row), which is used for output (4th row). 
    (c)~The low-pass \methodCut filter converts data into the frequency domain (2nd row), zeros out high-frequency components (3rd row), and converts the smoothed result back into the spatial domain (4th row). (d)~The \methodDP method subsamples the input data by iteratively selecting (2nd and 3rd rows) the points with the largest error to insert into the smoothed output (4th row).  All techniques are colored by type (see \autoref{fig:teaser}).}
    \label{fig:filter_illustrations}

\end{figure*}

\section{Taxonomy of Line Chart Smoothing Approaches}

We discuss 4 classes of smoothing that can be used on line charts. They can be broadly broken down into methods that consider local neighborhoods of data or global structures when determining the output. A summary of the 12 smoothing techniques analyzed in this paper can be found in \autoref{table:filters}. Each technique preserves some particular properties of the input through 1 or more adjustable simplification parameters. The number of available smoothing techniques is large. Therefore, this list is intended to be representative of well-known techniques, not necessarily comprehensive.

\subsection{Local Methods}

Local methods only consider nearby data when calculating their smoothed output. Essentially, for each output data, a local neighborhood of the input data is extracted. Then, the neighborhood is processed by a filter, and the result is used as the output.

\subsubsection{Rank Filters}

Rank filters are nonlinear filters that, for each input point, ranks (i.e., sorts) a neighborhood window surrounding the input point. A single value is selected from the ranked set for output. 
The \methodMed filter (see \autoref{fig:teaser:rank}(left)) selects the median value from the ranked neighborhood. The level of smoothing can be increased or decreased by enlarging or shrinking the neighborhood window, respectively. 
\methodMed filters are known for being particularly good at removing salt-and-pepper noise~\cite{arce2005nonlinear}, but if, on the other hand, those peaks represent important data, they will be lost with a \methodMed filter. 

To compute the \methodMed filter (see \autoref{fig:filter_illustrations:median}), for each of $n$ input points, a window of size $w$ is first selected. Next, the window is sorted. Finally, the median value is selected for output. The boundary of the domain requires special consideration. Several options exist for the boundary---we chose to repeat the boundary value infinitely. In a naive implementation of the \methodMed filter, repeated sorting operations are required, 1 per input/output point, making the overall performance $\BigO(n \cdot w \log{} w)$. The operation can be optimized by using a sliding window to achieve $\BigO(n \log{} w)$ in the general case~\cite{gil1993computing} and $\BigO(n)$ in limited cases~\cite{perreault2007median}. 

Additional examples of rank filters include \methodMin filter (see \autoref{fig:teaser:rank}(middle)) and \methodMax filter (see \autoref{fig:teaser:rank}(right)), which operate similarly, except that they select the minimum and maximum value from the ranked lists, respectively.

\subsubsection{Convolutional Filters}

Convolutional filters are a stencil-based method, where for a given input point, a series of weights are applied to a neighborhood surrounding that point. To compute a convolutional filter (see \autoref{fig:filter_illustrations:gaussian}), for each of $n$ input points, a window of size $w$ is selected. Next, the elements are multiplied by their corresponding elements from the stencil, summed, and that value is placed in the output. Similar to rank filters, the boundary of the domain requires special consideration. For consistency, we chose to repeat the boundary values infinitely. The resulting computational complexity for general convolutional filters is $\BigO( n \cdot w )$.

The \methodGau filter (see \autoref{fig:teaser:conv}(left)) is commonly used in convolutional signal and image processing~\cite{gaussian}. It weights the input neighborhood using a normal distribution. The smoothing level is increased or decreased by adjusting the standard deviation, $\sigma$, of the distribution. The \methodGau filter can be seen as a form of a low-pass filter, blurring both signal and noise from the data, producing smooth, visually appealing results. The window used for the \methodGau filter is fixed using $\sigma$ as a guide. In our implementation, a window size of $\pm 4 \sigma$ ensures that we capture over $99.9\%$ of the distribution. 

Another simple convolutional filter is the \methodMean filter (see \autoref{fig:teaser:conv}(right)), also known as the moving average. In this case, equal weights are applied to all elements in the window, resulting in the average being calculated. Because of the equal weighting, a sliding window can be used to improve performance to $\BigO( n )$ complexity.

Finally, \methodSG~\cite{savitzky1964smoothing} (see \autoref{fig:teaser:conv}(middle)) is a convolutional filter that uses a low-degree polynomial to smooth the data.

\begin{table}[!b]
\centering
\caption{Summary of Smoothing Algorithms Analyzed.}
\label{table:filters}
\resizebox{0.9\linewidth}{!}{%
\begin{tabular}{lll}
Technique    & Class         & Average Complexity                     \\
\hline
\methodMed   & Rank          & \hspace{7pt} $\BigO( n \log{} w )$     \\
\methodMin   & Rank          & \hspace{7pt} $\BigO( n )$              \\
\methodMax   & Rank          & \hspace{7pt} $\BigO( n )$              \\
\methodGau   & Convolutional & \hspace{7pt} $\BigO( n \cdot \sigma )$ \\
\methodMean  & Convolutional & \hspace{7pt} $\BigO( n )$              \\     
\methodSG    & Convolutional & \hspace{7pt} $\BigO( n \cdot w )$      \\
\methodCut   & Frequency     & \hspace{7pt} $\BigO( n \log{} n )$     \\
\methodButt  & Frequency     & \hspace{7pt} $\BigO( n \log{} n )$     \\
\methodCheb  & Frequency     & \hspace{7pt} $\BigO( n \log{} n )$     \\
\methodUni   & Subsampling   & \hspace{7pt} $\BigO( n + s )$              \\
\methodDP    & Subsampling   & \hspace{7pt} $\BigO( s \log{} n )$     \\
\methodTDA   & Subsampling   & \hspace{7pt} $\BigO(n + c \log{} c )$  \\
\end{tabular}
}\\
\vspace{2pt}
\begin{minipage}[t]{0.85\linewidth}\scriptsize $n$: number of input points; $w$: window size; $\sigma$: standard deviation of a normal distribution; $s$: number of output samples; $c$: number of critical points (i.e., local minimum or maximum)\end{minipage}
\end{table}

\subsection{Global Methods}

With global methods, the entire input data is considered in the calculation of the output.

\subsubsection{Frequency Domain Filters}
Frequency domain filtering converts the scalar data into a frequency domain representation, via wavelets or Fourier transform. Once in the frequency domain, undesirable frequencies are removed, and the signal is reconstructed. We consider a low-pass \methodCut filter (see \autoref{fig:teaser:freq}(left) and \autoref{fig:filter_illustrations:fft}), which converts the input into the frequency domain using a Discrete Fourier Transform (DFT)~\cite{cooley1965algorithm}. High-frequency components are then zeroed out to smooth the output above a cutoff frequency. Lowering that cutoff frequency increases the level of smoothing. Finally, the output is computed by converting the frequency domain data back to the spatial domain using an inverse DFT. Much like the \methodGau filter, the \methodCut filter produces smooth, visually appealing output, in this case, only retaining the specified frequencies. However, the relationship between the frequency and spatial domains is often not intuitive, as multiple frequencies contribute to a single output. The computational complexity of the DFT and the \methodCut filter is $\BigO( n \log{} n )$.

Additional frequency domain low-pass filters we consider include the \methodButt filter~\cite{butterworth1930theory} (see \autoref{fig:teaser:freq}(middle)) and \methodCheb filter~\cite{rhodes1980generalized} (see \autoref{fig:teaser:freq}(right)). The \methodCut filter is an idealized function that cannot be implemented in an electronic circuit, while the \methodButt and \methodCheb filters can. Practically speaking, these methods differ from the \methodCut filter in that they provide a gradual ramp-down of the cutoff frequency.

\begin{figure*}[!t]
    \centering
    \begin{minipage}[b]{0.77\textwidth}
        \subfigure[Total Variation\label{fig:measures:variation}]{\includegraphics[width=0.32\linewidth]{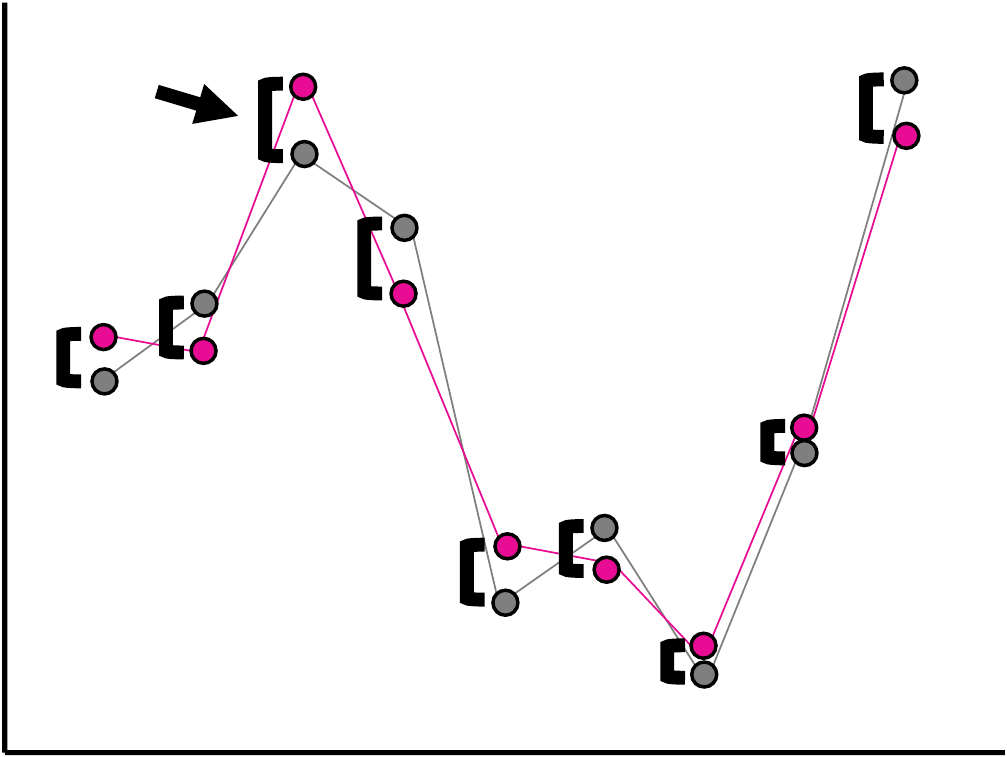}}
        \hfill
        \subfigure[Area Preservation\label{fig:measures:volume}]{\includegraphics[width=0.32\linewidth]{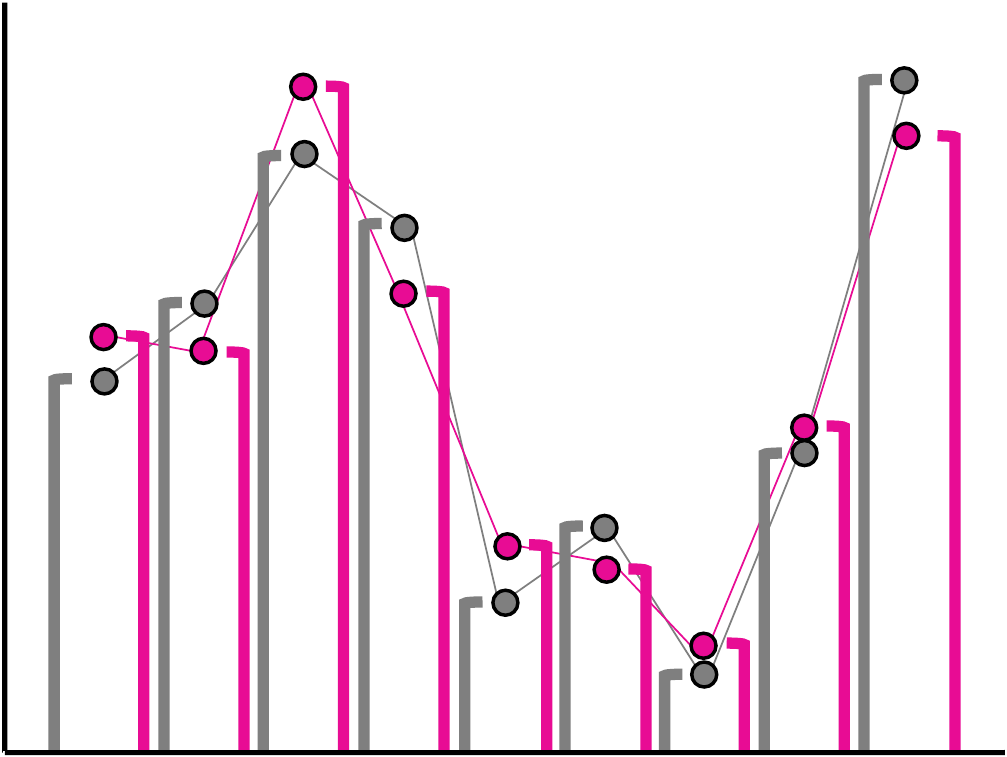}}
        \hfill
        \subfigure[Peaks\label{fig:measures:peaks}]{\includegraphics[width=0.32\linewidth]{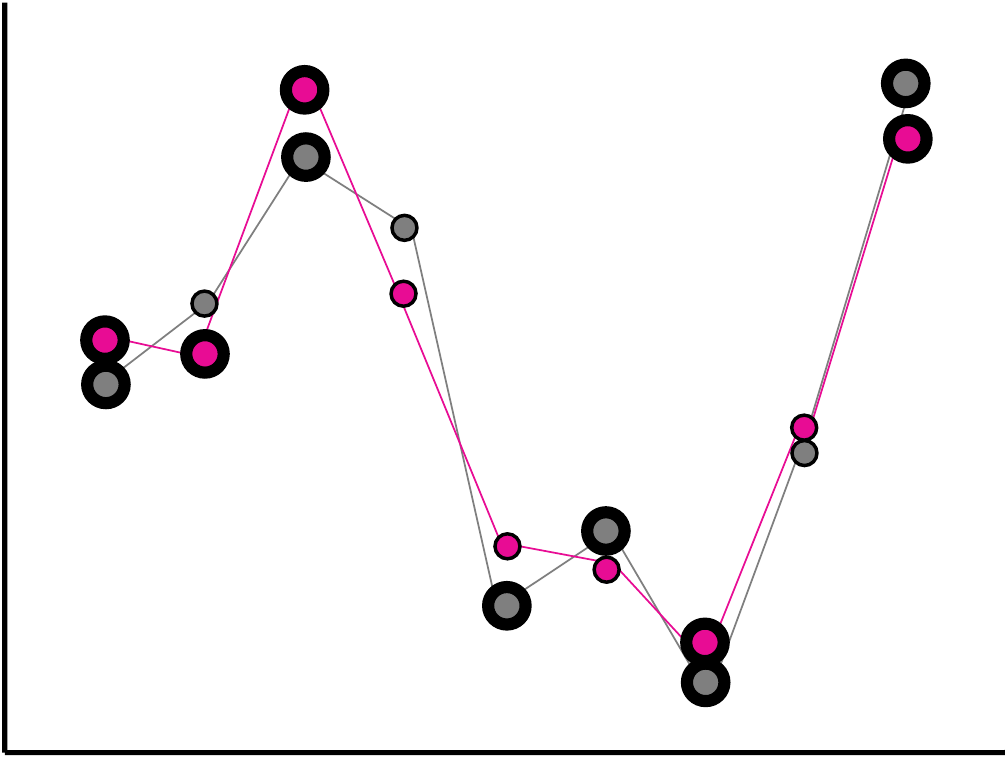}}
    \end{minipage}
    \begin{minipage}[b]{0.225\textwidth}
        \centering
        \subfigure[Frequency Preservation\label{fig:measures:freq}]{\includegraphics[width=0.99\linewidth]{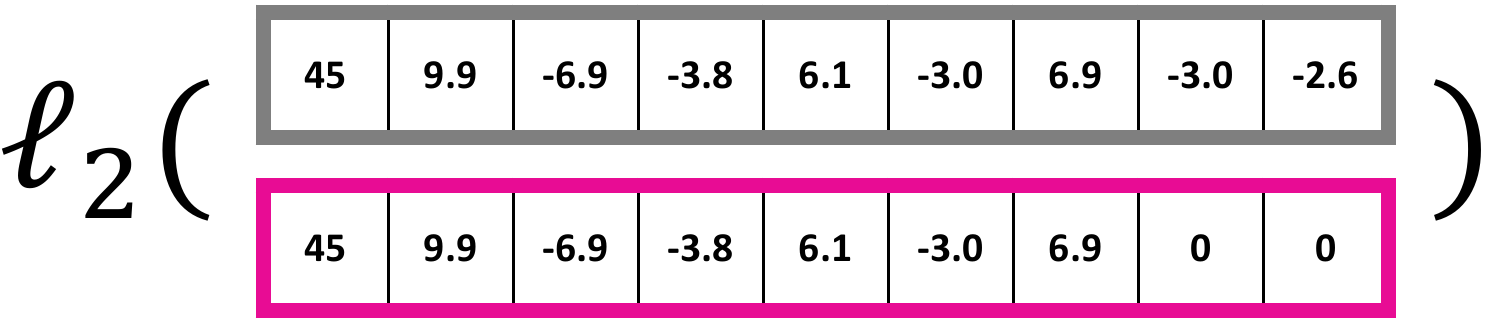}}
        \subfigure[Value-Order Preservation\label{fig:measures:order}]{\hspace{10pt}\includegraphics[width=0.9\linewidth]{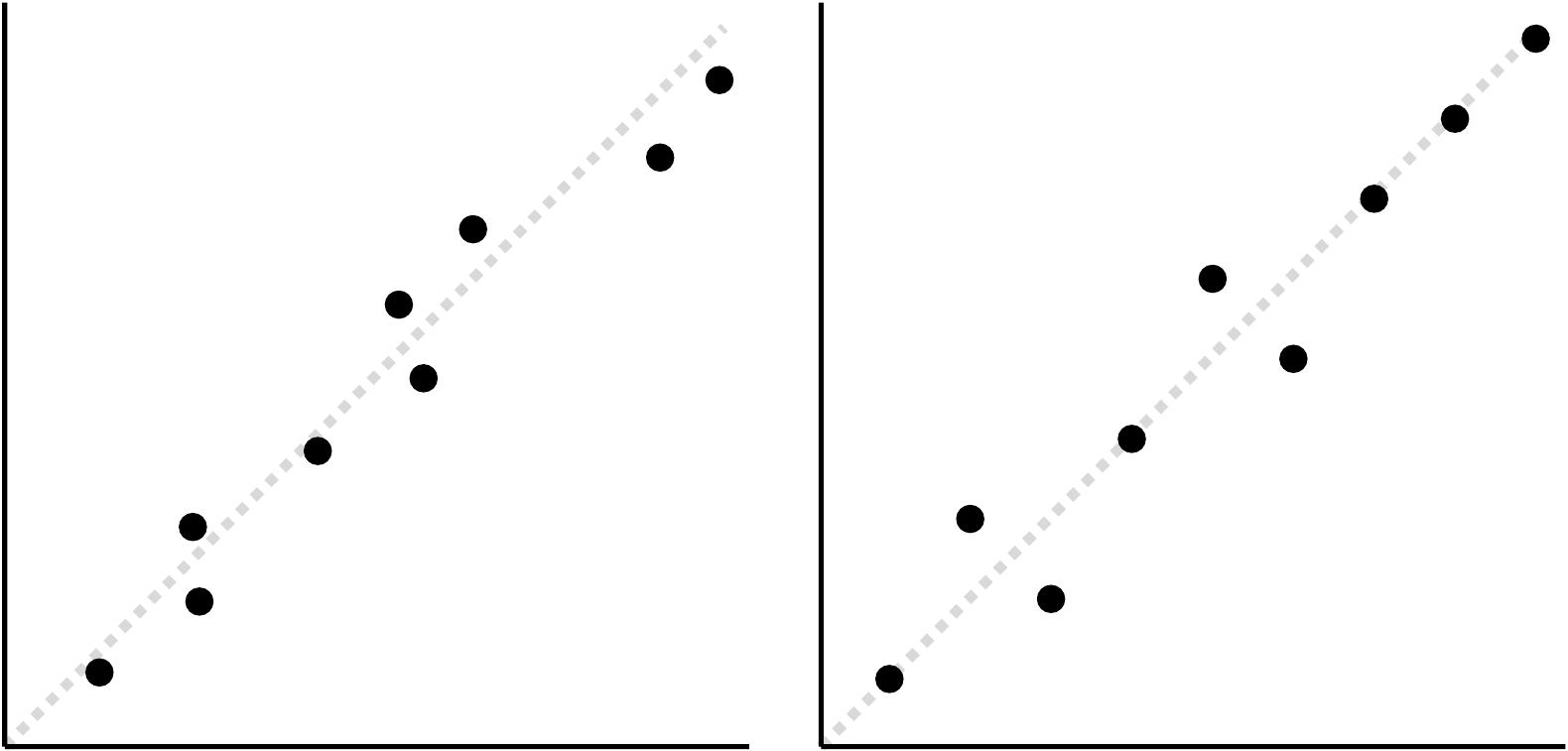}\hspace{10pt}}
    \end{minipage}
    \caption{Illustration of the 5 types of effectiveness measurement used in our approach. The example is based upon the low-pass \methodCut filter from \autoref{fig:filter_illustrations:fft}. (a)~shows the total variation, where $\lnorm_{1}$ is the sum of the black brackets, and $\lnorm_{\infty}$ is the value of the largest (see arrow). (b)~measures the change in the area, $\delta a$, by taking the difference between the sum of the grey bars and the sum of the pink bars. (c)~illustrates the peaks in the data in bold. The $W_{1}$ measures the total difference between peaks, while $W_{\infty}$ measures the largest variation. (d)~shows the frequency domain of the input and smoothed data. $\Fgroup$ measures the $L^2$-norm of the difference between these 2 vectors. (e) illustrates the (left) value preservation, $\rho$, and (right) order preservation, $r_s$, of the smoothed line chart. Proximity to the diagonal indicates how well the value/order is preserved.}
    \label{fig:measures}
\end{figure*}

\subsubsection{Subsampling}

Subsampling approaches take the original data and select a subset of the original data points as representatives of the whole data. Simplification is increased by merely selecting fewer points. 

A common choice, due to its ease of implementation, \methodUni subsampling (see \autoref{fig:teaser:subs}(left)) selects points at regular intervals. Between selected points, interpolation is used, with linear interpolation being the most straightforward case. \methodUni subsampling makes few guarantees about the types of features it preserves unless the input is already \textit{oversampled}, in which case it retains the original signal~\cite{shannon1949communication}. Computationally, \methodUni subsampling is very efficient, only $\BigO(n+s)$, where $s$ is the number of samples taken from the input.

Nonuniform subsampling, in contrast to \methodUni subsampling, selects points at irregular intervals by considering/preserving some features of the data. \methodDP~\cite{ramer1972iterative,douglas1973algorithms} (see \autoref{fig:teaser:subs}(middle)) is an example that establishes a priority queue of points by optimizing the $L^\infty$-norm of the residual error (i.e., the difference between the original and smoothed line charts). The algorithm (see \autoref{fig:filter_illustrations:dpr}) starts by selecting the boundary points of the input data (i.e., first and last points) for initialization and connects them via linear interpolation. Points are then iteratively added by selecting the input point with the largest distance from the current output and inserting it into the output. The process continues until a user-specified threshold distance is reached. The simplification is increased or decreased by modifying this threshold. The output captured by \methodDP is reliable and predictable, in that the output will deviate no more than the specified threshold. The worst-case complexity of the algorithm is $\BigO(n^2)$, while the average complexity is $\BigO(n \log{} n)$.

An additional nonuniform subsampling approach, the \methodTDA filter (see \autoref{fig:teaser:subs}(right)), uses techniques from Topological Data Analysis to smooth data in a way that retains significant peaks and minimizes error~\cite{suh2019topolines}. The \methodTDA filter works by first identifying critical points, in the form of local minima and local maxima, and forms a hierarchical pairing (1 each, a local minimum and local maximum) between them. Pairs of critical points are then removed from the output if the difference in their value is below a given simplification threshold. Finally, monotonic regression is used to interpolate between the remaining critical points. The overall complexity of the operation is $\BigO( n + c \log{} c )$, where $c$ is the number of local minima and maxima.

%% file: sec-framework-metrics.tex
\section{Analytical Framework for Measuring for Smoothing Efficacy in Line Charts}

In this section, we describe our analytical framework for evaluating line chart smoothing. It consists of 3 parts: a set of effectiveness measures for line chart smoothing (see \autoref{sec:measures}); a description of the relationship between the effectiveness measures and common visual analytics tasks (see \autoref{sec:tasks}); and a description of the methodology for comparing different line chart smoothing techniques (see \autoref{sec:eval:framework}).

\subsection{Measures of Effectiveness}
\label{sec:measures}

To better understand the \textit{quality} of smoothing results produced by each smoothing technique, we consider a set of measures that compare the input data, $X =\{x_0,x_1,..,x_i,..,x_n\}$, and the smoothed data, $Y =\{y_0,y_1,..,y_i,..,y_n\}$. There is no single measure to evaluate the effectiveness of smoothing under all visual analytics tasks. Therefore, we use a series of measures, each of which relates how well the smoothing technique preserves a particular quality of the input data. For all measures, a value of 0 indicated no error, while larger positive values indicate increasing errors.

\subsubsection{Total/Maximum Value Variation}
\label{sec:measures:value_var}

The first measures consider calculating the difference between the input and the smoothed data using vector norms, which measure and sum the difference between the data at each sample location. Considering the illustration in \autoref{fig:measures:variation}, we apply 2 variations, the $L^1$-norm and the $L^\infty$-norm.

The $L^1$-norm, $\lnorm_1$, also known as the least absolute deviations or least absolute errors, measures the sum of the absolute value of the difference between the input and smoothed data. In other words, in \autoref{fig:measures:variation}, it measures the sum of the differences in black. As a measure, it is robust, in that it is resistant to the influence of outliers. The $L^1$-norm is:
\begin{equation}
    \lnorm_{1}(X,Y)=\sum _{i=1}^{n}\left|x_{i}-y_{i}\right|
    \label{eqn:l1}
\end{equation}

The $L^\infty$-norm, $\lnorm_{\infty}$, measures only the point of the largest difference between input and smoothed data. In \autoref{fig:measures:variation}, this is the point denoted by the arrow. The $L^\infty$-norm is: 
\begin{equation}
    \lnorm_{\infty}(X,Y)=\max\limits_i \left|x_{i}-y_{i}\right|
    \label{eqn:linf}
\end{equation}

\subsubsection{Area Preservation}
\label{sec:measures:volume}
In some cases, the individual deviations matter less than the total area captured under the line chart. The change in the area, $\delta a$, is found by taking the difference between the integrals of the input and smoothed data. \autoref{fig:measures:volume} illustrates the process. The change in area is the difference between the sum of all grey bars and the sum of all pink bars. The change in area is:
\begin{equation}
    \delta a(X,Y)=\left|\sum _{i=1}^{n} x_{i}- \sum _{i=1}^{n} y_{i}\right|
    \label{eqn:vol}
\end{equation}

\subsubsection{Total/Maximum Peak Variation}
\label{sec:measures:peaks}
The next measure identifies and matches the similarity of peaks, i.e., local minima and maxima, between the original and smoothed data. \autoref{fig:measures:peaks} shows examples of such peaks. 
To measure the similarity, we use techniques from Topological Data Analysis~\cite{EdelsbrunnerHarer2010}. 
First, the local minima and maxima of the original and smoothed data are calculated and paired in a process described in detail in~\cite{suh2019topolines}. The pairs are placed into 2 sets $\overline{X}$ and $\overline{Y}$\footnote{For technical reasons, all diagonal points $(\overline{x},\overline{x})$ are added to make the cardinality infinite~\cite{kerber2017geometry}.}, and let $\eta$ be a bijection between the two sets.

The Wasserstein distance measures the total difference between all peaks, giving higher weight to those with larger differences. The 1-Wasserstein distance, $W_1$, is:
\begin{equation}
    W_1(\overline{X},\overline{Y}) = \displaystyle \inf_{\eta:\overline{X} \rightarrow \overline{Y}}  \sum_{\overline{x}\in \overline{X}} \left\lVert \overline{x}-\eta(\overline{x}) \right\rVert_1
    \label{eqn:wass}
\end{equation}

The Bottleneck distance only measures the peaks with the maximum difference. The Bottleneck distance is:
\begin{equation}
    W_{\infty}(\overline{X},\overline{Y}) = \inf_{\eta: \overline{X} \rightarrow \overline{Y}} \sup_{\overline{x} \in \overline{X}} \left\lVert \overline{x}-\eta(\overline{x}) \right\rVert_\infty 
    \label{eqn:bottleneck}
\end{equation}

\subsubsection{Frequency Preservation}
\label{sec:measures:freq}
Generally speaking, a smoothed signal should maintain as much of the frequency spectrum as possible. To measure the preservation of frequencies, $\Fgroup$, we convert the original and smoothed data into the frequency domain using the Discrete Fourier Transform (DFT), $F_X$ and $F_Y$, respectively. Once the DFTs are calculated, their difference is found using the $L^2$-norm between them:
\begin{equation}
    \Fgroup(F_X,F_Y)=\sqrt{\sum _{k=1}^{n}\left(F_{X,k}-F_{Y,k}\right)^2},
    \label{eqn:freqp}
\end{equation}
where $k$ is a single frequency of interest. \autoref{fig:measures:freq} illustrates the frequency domain before and after smoothing. The frequency preservation would be the $L^2$-norm of the difference between these 2 vectors.

\subsubsection{Value-Order Preservation}
\label{sec:measures:order}
In some scenarios, knowing that the relative values of data items are maintained is more important than maintaining the correct values. The value-order relationship can be measured using the correlation between the input and smoothed data. To measure the relationship between relative values, the Pearson Correlation Coefficient, $\rho$, can be employed. \autoref{fig:measures:order}(left) illustrates the value relationship by placing points at $(x_i,y_i)$. Intuitively, the $\rho$ measures the proximity of the points to the diagonal in grey. In order to treat $\rho$ consistently with other measures, we modify it, such that 0 is a perfect positive correlation, and 2 is a perfect negative correlation. The modified $\rho$ is:
\begin{equation}
    \rho(X,Y)=1-\frac{\operatorname{cov}(X,Y)}{\sigma_{X}\sigma_{Y}}
    \label{eqn:pcc}
\end{equation}

The order relationship between data items can be measured using Spearman Rank Correlation, $r_s$, which is the Pearson Correlation Coefficient of the ranked data, in other words:
\begin{equation}
    r_s(X,Y) = \rho(rank(X),rank(Y))
    \label{eqn:rs}
\end{equation}
\autoref{fig:measures:order}(right) illustrates the order relationship. The points are placed by rank, instead of value, and their $\rho$ is calculated.

%% file: sec-framework-tasks.tex
\subsection{Low-Level Task Taxonomy for Line Charts}
\label{sec:tasks}

To determine relevant visual analytics tasks, we adapt the low-level task taxonomy of Amar et al.~\cite{amar2005low} to line charts. For each task, we provide a brief description and an example query. Finally, we relate each of these tasks to 1 or 2 of the metrics from the previous section, which is summarized in \autoref{tbl:tasks}.

\newcommand{\CM}{\multicolumn{2}{c|}{\checkmark}}
\newcommand{\EM}{&}
\newcommand{\mrX}{\multirow{2}{*}{X}}
\newcommand{\theaderA}[1]{\multicolumn{2}{c}{
    \begin{minipage}[b]{23pt}
        \hspace{-100pt}
        {\begin{minipage}[b]{0pt}\rotatebox[origin=l]{-25}{{\begin{minipage}[b]{125pt}\begin{flushright}\large #1\end{flushright}\end{minipage}}}\end{minipage}}
        \end{minipage}}}

\newcommand{\theaderB}[2]{\multicolumn{2}{c}{
    \begin{minipage}[b]{23pt}
        \hspace{-95pt}
        {\begin{minipage}[b]{0pt}\rotatebox[origin=l]{-35}{\begin{minipage}[b]{115pt}\begin{flushright}#2\end{flushright}\end{minipage}}\end{minipage}}
        \hspace{12pt}
        {\begin{minipage}[b]{0pt}\rotatebox[origin=l]{-35}{{\begin{minipage}[b]{115pt}\begin{flushright}#1\end{flushright}\end{minipage}}}\end{minipage}}
        \end{minipage}}}
        
\newcommand{\rowHeaderA}[2]{
\multicolumn{2}{r|}{\begin{minipage}[c][18pt][c]{100pt}\flushright #1\\#2\end{minipage}}
}
\newcommand{\rowHeaderB}[1]{
\multicolumn{2}{r|}{\begin{minipage}[c][25pt][c]{100pt}\flushright\large #1\end{minipage}}
}

\renewcommand*{\arraystretch}{0.5}
\begin{table}[!t]
    \centering
    \caption{Matrix of Tasks and Metrics}
    \label{tbl:tasks}
    
    \resizebox{0.95\linewidth}{!}{%
\begin{tabular}{@{}r@{}r|@{}c@{}c@{}|@{}c@{}c@{}|c@{}c|c@{}c|c@{}c|c@{}c|cc|cc|cc|}
   \multicolumn{2}{c}{}
 & \theaderA{Total Value Variation ($\lnorm_{1}$)} 
 & \theaderA{Maximum Value Variation ($\lnorm_{\infty}$)} 
 & \theaderA{Area Preservation ($\delta a$)}  
 & \theaderA{Total Peak Variation ($W_1$)}
 & \theaderA{Maximum Peak Variation ($W_{\infty}$)} 
 & \theaderA{Frequency Preservation ($\Fgroup$)}
 & \theaderA{Value Preservation ($\rho$)}
 & \theaderA{Order Preservation ($r_s$)} \\
 
\rowHeaderB{Retrieve Value}            & \CM & \CM & \EM & \EM & \EM & \EM & \EM & \EM \\
\hline
\rowHeaderB{Determine Range}           & \CM & \CM & \EM & \EM & \EM & \EM & \EM & \EM \\
\hline
\rowHeaderB{Compute Derived Value}     & \EM & \EM & \CM & \EM & \EM & \EM & \EM & \EM \\
\hline
\rowHeaderB{Find Extrema}              & \EM & \EM & \EM & \CM & \CM & \EM & \EM & \EM \\
\hline
\rowHeaderB{Find Anomalies}            & \EM & \EM & \EM & \CM & \CM & \EM & \EM & \EM \\
\hline
\rowHeaderB{Characterize Distribution} & \EM & \EM & \EM & \EM & \EM & \CM & \EM & \EM \\
\hline
\rowHeaderB{Sort}                      & \EM & \EM & \EM & \EM & \EM & \EM & \CM & \CM \\ 
\hline
\multirow{2}{*}{\hspace{45pt}\large Cluster} & \footnotesize Trends & \EM & \EM & \EM & \EM & \EM & \CM & \EM & \EM \\ 
                                            & \footnotesize Points & \EM & \EM & \EM & \EM & \EM & \EM & \CM & \CM \\
\hline
\end{tabular}
}
\end{table}

\begin{figure}[!b]
    \centering
    \includegraphics[width=0.75\linewidth]{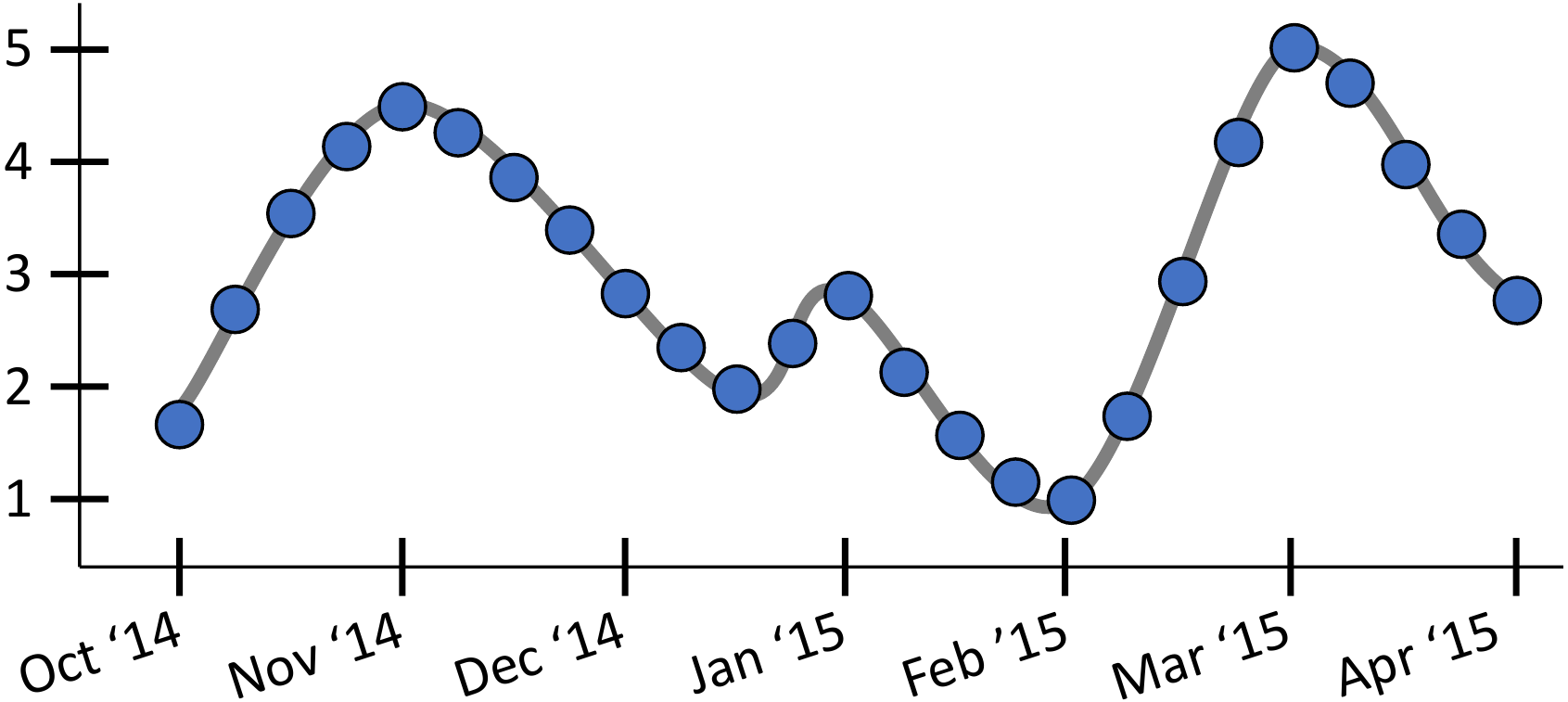}
    \hspace{5pt}
    \caption{Example ``stock price'' line chart for demonstrating task queries.}
    \label{fig:task_ex}
\end{figure}

\newcommand{\keepvalues}{%
  \edef\restorevalues{%
    \intextsep=\the\intextsep
    \columnsep=\the\columnsep
  }%
}

\keepvalues

\setlength{\intextsep}{2pt}%
\setlength{\columnsep}{4pt}%

\vspace{3pt}
\begin{wrapfigure}[5]{L}{0.435\linewidth}
    \includegraphics[width=1\linewidth]{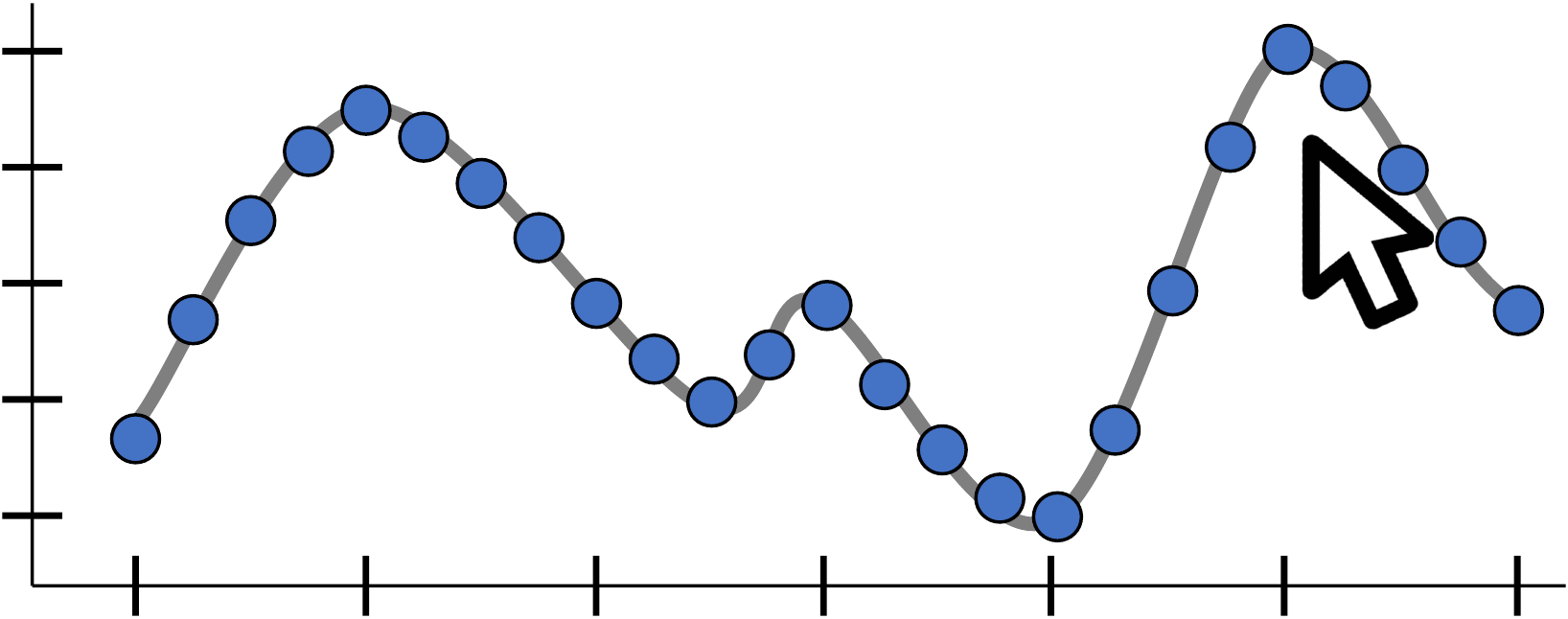}
\end{wrapfigure}
\noindent The \textbf{Retrieve Value} task is focused on finding the function value at an exact location in a given dataset/chart. For example, using \autoref{fig:task_ex}, ``What is the stock price on Mar~'15?'' (answer: $\sim4.5$). The accuracy of retrieving a value is dependent upon how closely chart values match the data value, which mirrors the total/maximum value variation measures (see \autoref{sec:measures:value_var} and \autoref{fig:measures:variation}). Measuring the total difference between the smoothed data and the original data, in other words, the $L^1$-norm (\autoref{eqn:l1}), provides an \textit{average case} performance for the retrieve value task. By measuring the maximum difference between the smoothed and original function, in other words, the $L^\infty$-norm (\autoref{eqn:linf}), the \textit{worst case} performance can be calculated.

\vspace{3pt}
\begin{wrapfigure}[5]{L}{0.435\linewidth}
    \includegraphics[width=\linewidth]{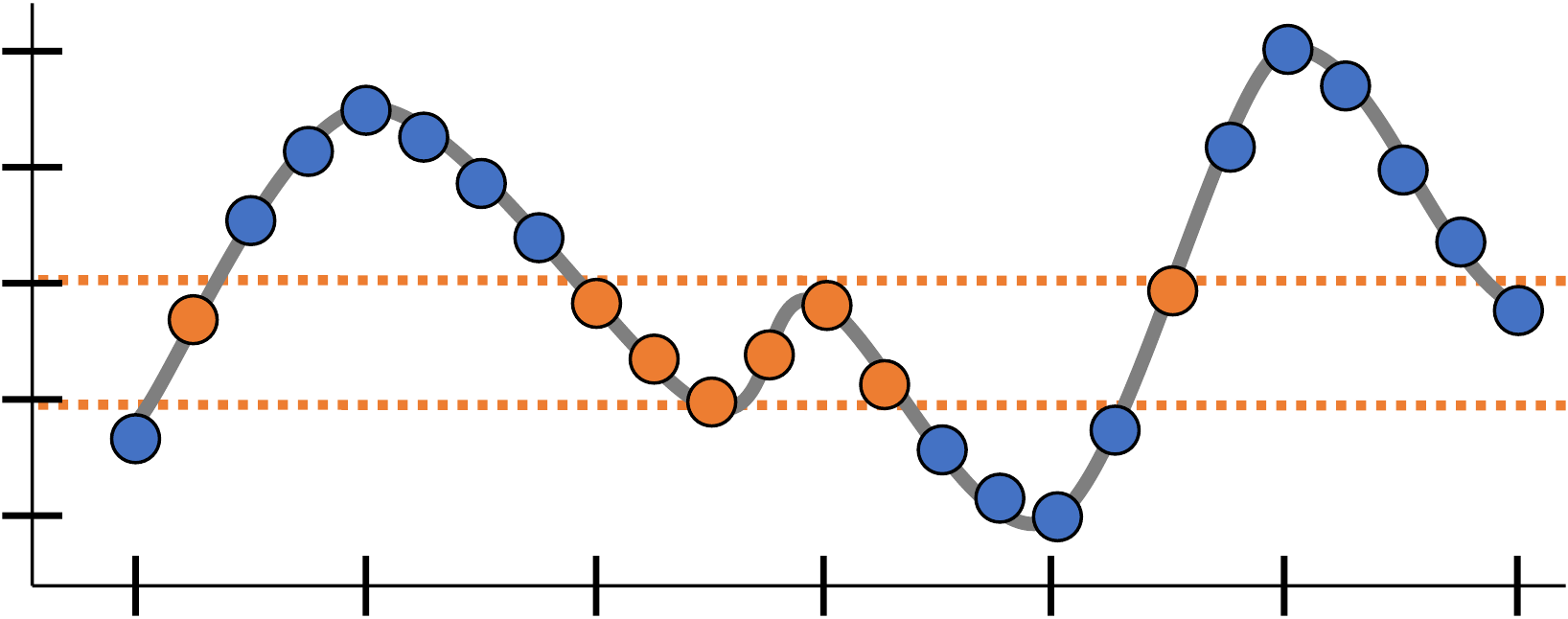}
\end{wrapfigure}
\noindent For the \textbf{Determine Range} task, specific criteria, e.g., a range of values, are provided for identifying data points, e.g., the dates/times that are within that range.  Using \autoref{fig:task_ex}, an example is, ``What months saw values between 3 and 4?'' (answer: Oct~'14, Nov~'14, Feb~'15, Mar~'15). Accuracy in performing this task is highly dependent upon the criteria provided. Nevertheless, generally, it is important that the values of the data closely reflect those of the input data, in other words, the total/maximum value variation (see \autoref{sec:measures:value_var} and \autoref{fig:measures:variation}). The \textit{average case} is measured using the $L^1$-norm (\autoref{eqn:l1}). The \textit{worst case} is measured using the $L^\infty$-norm (\autoref{eqn:linf}).

\vspace{3pt}
\begin{wrapfigure}[5]{L}{0.435\linewidth}
    \includegraphics[width=\linewidth]{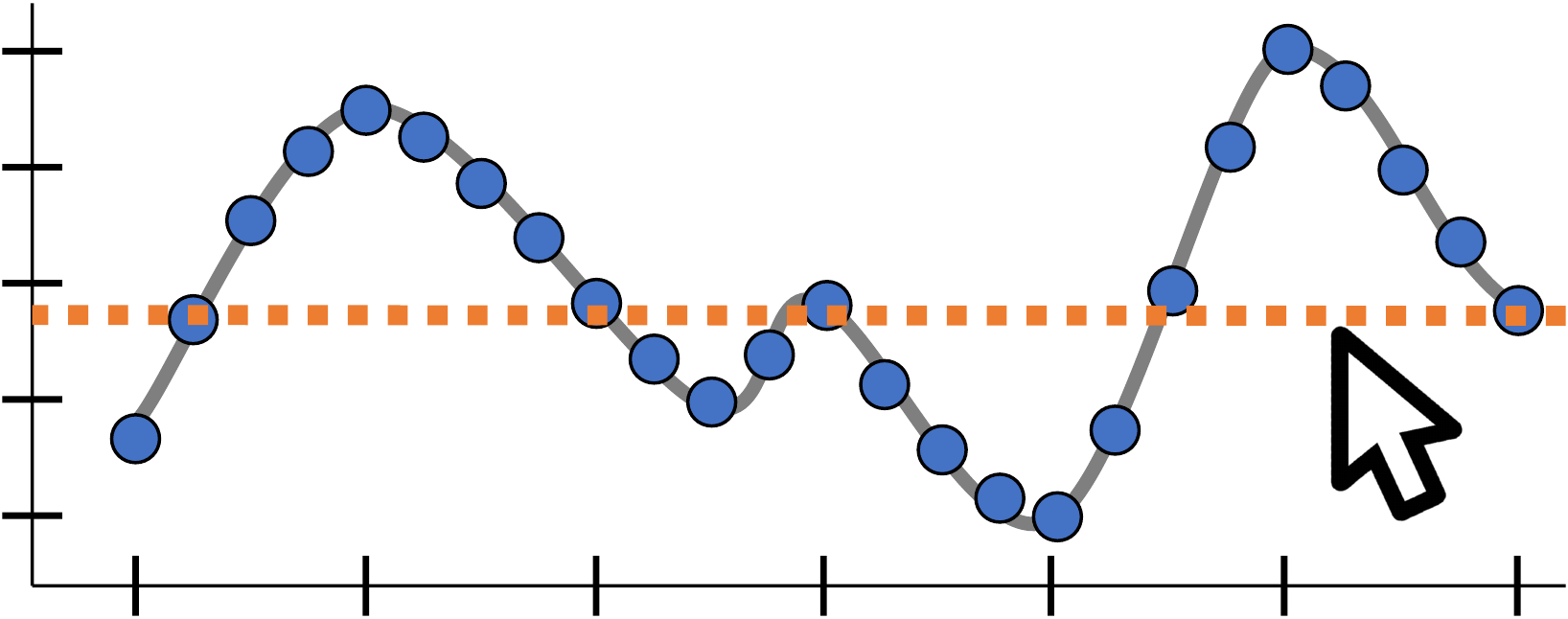}
\end{wrapfigure}
\noindent The \textbf{Compute Derived Value} task focuses on computing an aggregate, such as the average or total value of a function. For example, using \autoref{fig:task_ex}, ``What is the average stock price from Oct~'14 to Apr~'15?'' (answer: $\sim3$). The task accuracy is mostly dependent on how well the line chart globally (i.e., the sum of all values) matches the input data.
The task essentially requires the user to visually determine an integral of the data, which is equivalent to the area preservation measure (see \autoref{sec:measures:volume} and \autoref{fig:measures:volume}). The area preservation measure, $\delta a$, provides the \textit{average case} performance by measuring the difference in integrals between the original and smoothed data.

\vspace{3pt}
\begin{wrapfigure}[5]{L}{0.435\linewidth}
    \includegraphics[width=\linewidth]{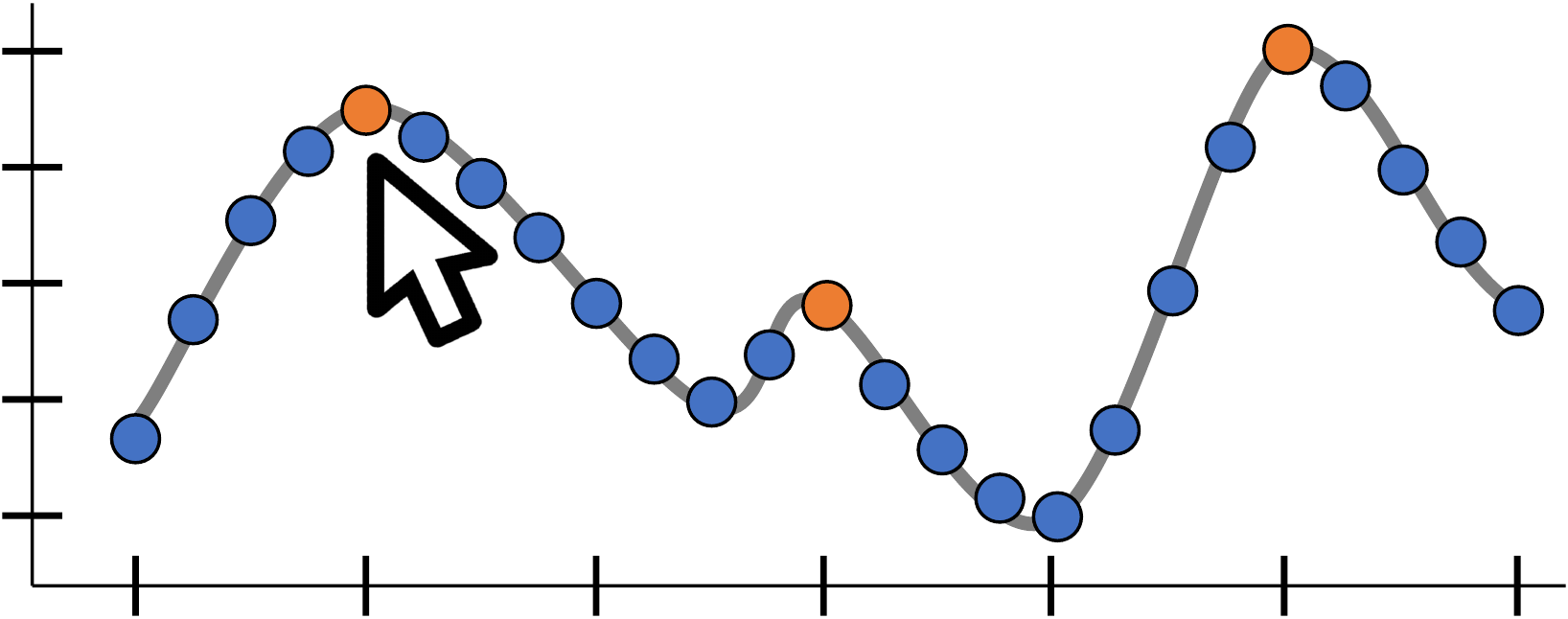}
\end{wrapfigure}
\noindent The \textbf{Find Extrema} task is concerned with finding local minima and maxima (i.e., valleys and peaks) in the data. For example, using \autoref{fig:task_ex}, ``What are the dates/values of all of the peaks in the data?'' (answer: Nov~'14/4.5, Jan~'15/3, Mar~'15/5). The accuracy of the task depends upon the peaks and/or valleys remaining present in the output and significant enough to be visible. The total/maximum peak variation (see \autoref{sec:measures:peaks} and \autoref{fig:measures:peaks}) measures both the existence and significance of peaks in the data. The \textit{average case} is provided by the 1-Wasserstein distance (\autoref{eqn:wass}), which finds the total variation in peaks between the input and smoothed data. The \textit{worst case} is found using the Bottleneck distance (\autoref{eqn:bottleneck}), which measures the peak of maximal variation.

\vspace{3pt}
\begin{wrapfigure}[5]{L}{0.435\linewidth}
    \includegraphics[width=\linewidth]{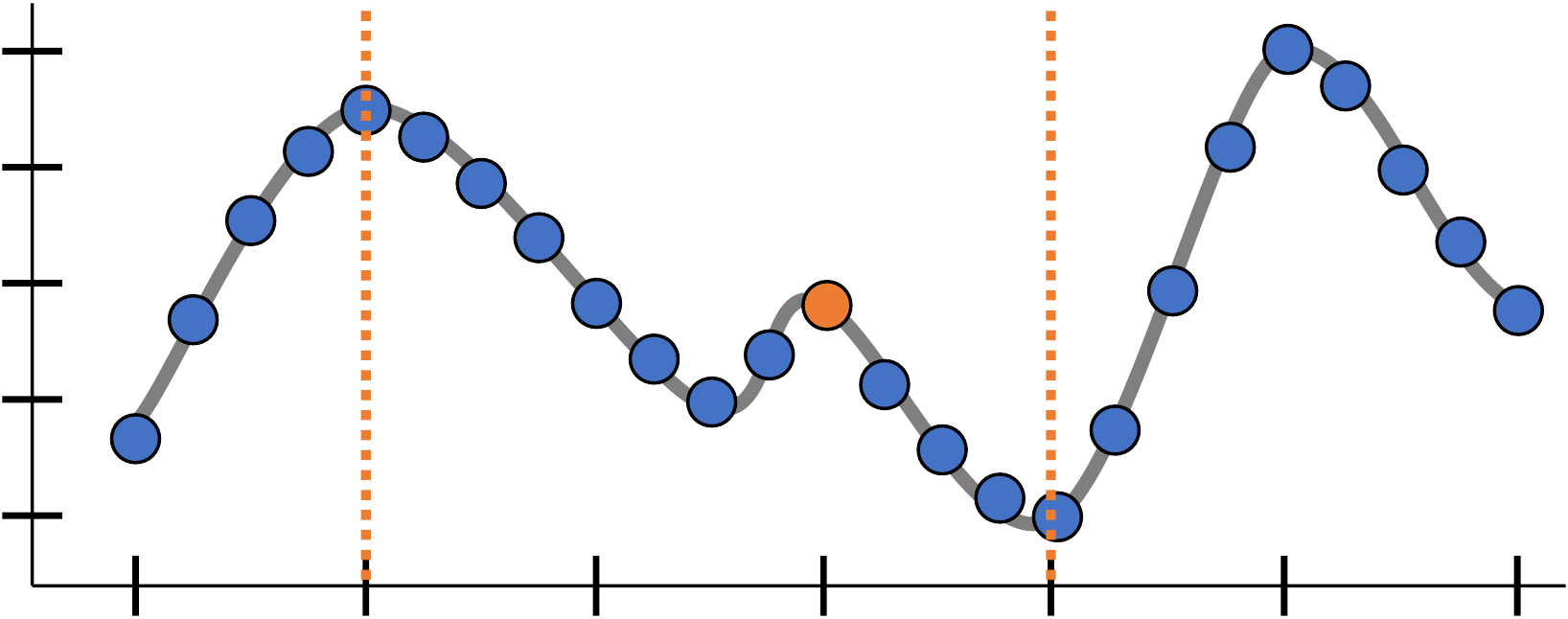}
\end{wrapfigure}
\noindent The \textbf{Find Anomalies} task involves looking for values that do not conform to the overall trend in the data. For example, ``Between Nov~'14 and Feb~'15, what month, if any, does not follow the data trend?'' (answer: Dec~'14/Jan~'15). The task of finding anomalies is similar to finding extrema, in that anomalies are generally peaks in the data, but in this case, they do not follow the trend of the data. Since the task involves identifying peaks, the \textit{average case} is provided by the 1-Wasserstein distance (\autoref{eqn:wass}), and the \textit{worst case} is found using the Bottleneck distance (\autoref{eqn:bottleneck}). Interestingly, the removal of anomalies is also one of the reasons smoothing is applied to line charts. Therefore, when performing other tasks, the \textit{preservation of anomalies might be considered a negative quality}.

 \vspace{3pt}
\begin{wrapfigure}[5]{L}{0.435\linewidth}
    \includegraphics[width=\linewidth]{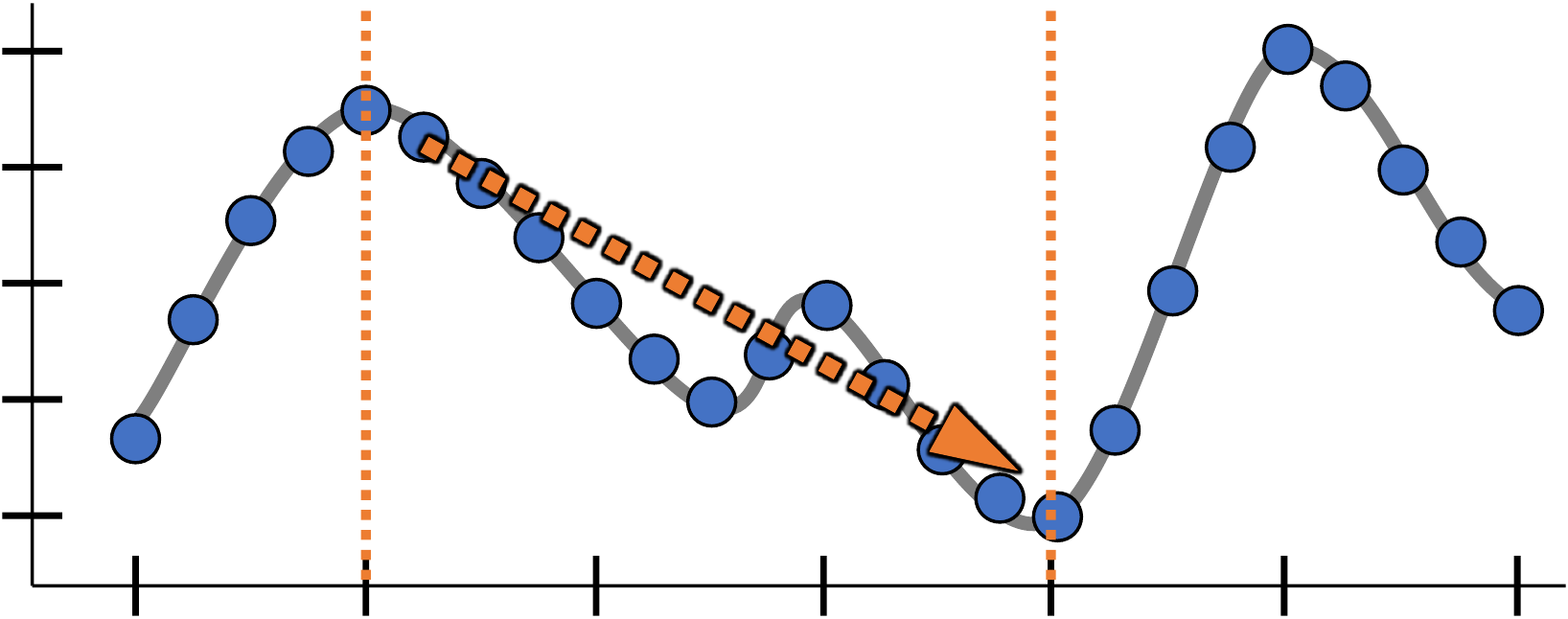}
\end{wrapfigure}
\noindent The \textbf{Characterize Distribution} task involves summarizing a trend in the data. For example, using \autoref{fig:task_ex}, ``What is the trend in the data between Nov~'14 and Feb~'15?'' (answer: downward). Trends in the data are synonymous with the frequency domain of the data. To be effective, the frequency domain of the smoothed data should be as similar as possible to that of the input data. Therefore, the \textit{average case} accuracy of this task is measurable using the frequency preservation measure (see \autoref{sec:measures:freq} and \autoref{fig:measures:freq}) from \autoref{eqn:freqp}.

\vspace{3pt}
\begin{wrapfigure}[5]{L}{0.435\linewidth}
    \includegraphics[width=\linewidth]{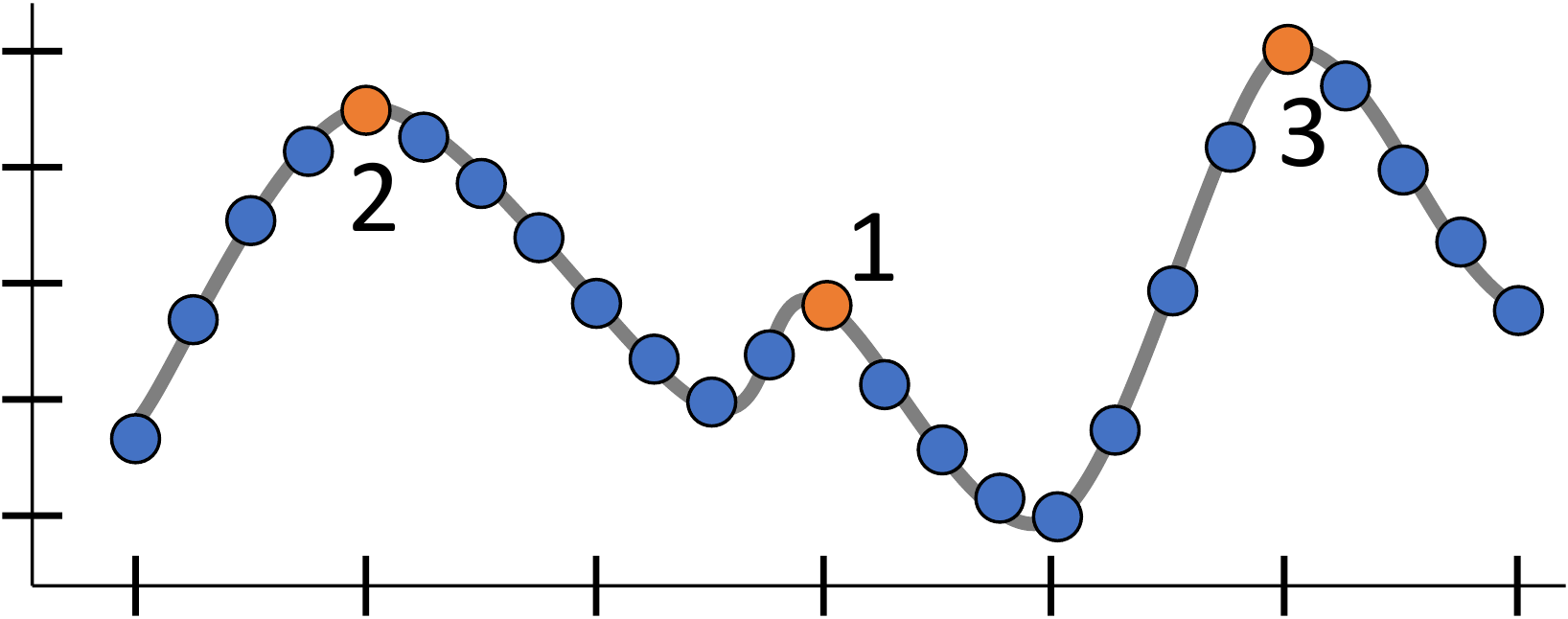}
\end{wrapfigure}
\noindent The \textbf{Sort} task asks users to, given some criteria, rank or order the data values. An example, using \autoref{fig:task_ex}, would be, ``What is the order of stock values for the dates Nov~'14, Jan~'15, and Mar~'15, from lowest to highest?'' (answer: Jan~'15, Nov~'14, Mar~'15). While similar to the retrieving value task, the accuracy of this task relies both upon the relative order of values (not the exact values) of data remaining the same, and further, the difference between those relative values is reasonably discernible. The value-order preservation measures (see \autoref{sec:measures:order} and \autoref{fig:measures:order}) provide 2 mechanisms to understand the \textit{average case} performance. First, Spearman Rank Correlation (see \autoref{eqn:rs}) can be used to compare the relative order of all points in the original and smoothed data. The Pearson Correlation Coefficient (see \autoref{eqn:pcc}) can be used to determine, on average, how discernible the values in the smoothed data are from one another, as compared to the input.

\vspace{3pt}
\begin{wrapfigure}[5]{L}{0.435\linewidth}
    \includegraphics[width=\linewidth]{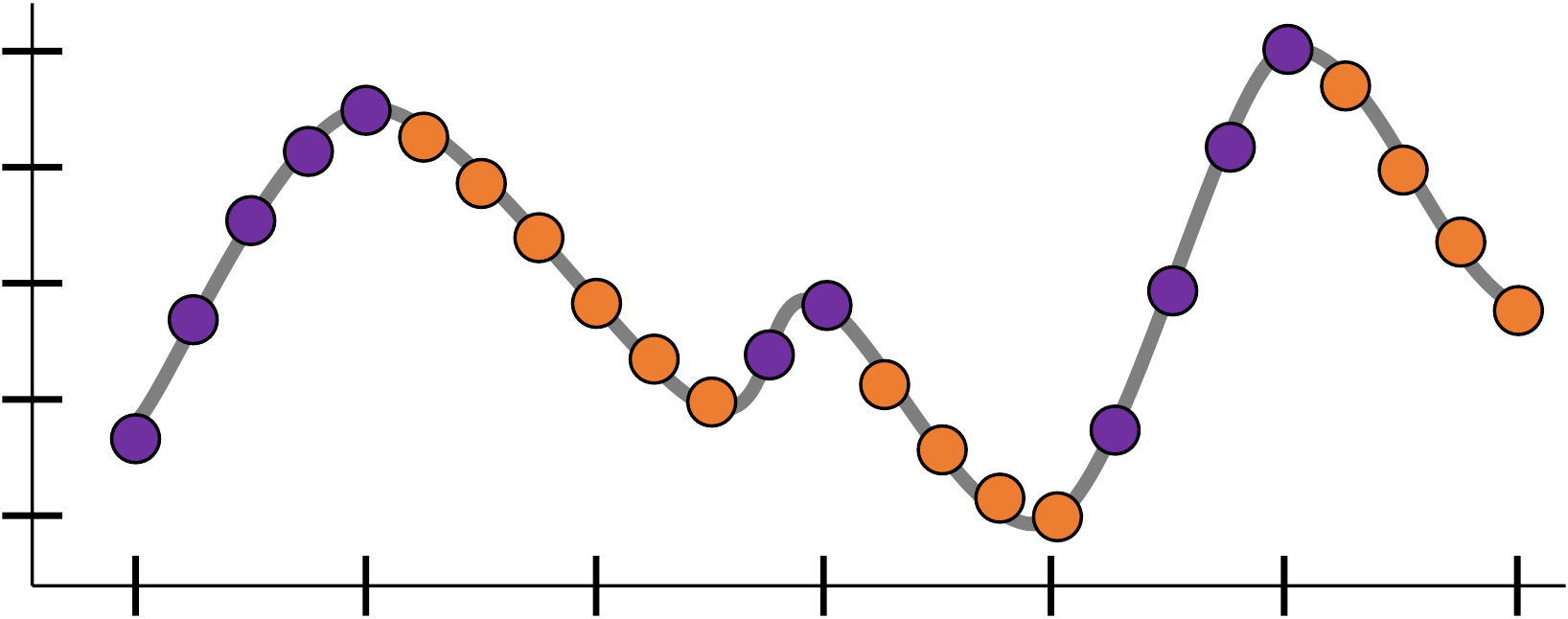}
\end{wrapfigure}
\noindent The \textbf{Cluster} task asks users to group data with similar values or trends. For example, ``Group months with similar trends'' (answer: see left). Depending upon the nature of the query (clustering individual points vs.\ clustering trends), this task depends upon both the judgment of relative values and trends in the data. Therefore, the \textit{average case} performance is summarized by the frequency preservation measure (see \autoref{sec:measures:freq} and \autoref{fig:measures:freq}) from \autoref{eqn:freqp} for clustering trends, as well as the value-order presentation measures (see \autoref{sec:measures:order} and \autoref{fig:measures:order}), Spearman Rank Correlation (see \autoref{eqn:rs}) and Pearson Correlation Coefficient (see \autoref{eqn:pcc}) for clustering individual points.

\restorevalues{}

\paragraph{Tasks Not Considered} Amar et al.~\cite{amar2005low} defined additional low-level tasks that we found redundant or out-of-scope for this analysis. First was the \textit{filter} task, which we felt was redundant with tasks such as determine range, characterize distribution, and find anomalies. The second was the \textit{correlate} task, which we felt would require comparing multiple distributions for their similarity. While this task could potentially be analyzed with our framework, we only consider the effectiveness of tasks on a single line chart.

%% file: sec-framework-eval.tex
\subsection{Evaluation Framework}
\label{sec:eval:framework}
Given the metrics and tasks described in the prior subsections, we establish our framework for comparing the efficacy of smoothing techniques. The idea is to rank the effectiveness of all smoothing techniques for given data on a specific visual analytics task.

This requires 2 things: (1) a common measure of the smoothing level (i.e., a baseline for measurement); and (2)~a method for ranking the efficacy of methods, using a specific metric.

\subsubsection{Visual Complexity as a Proxy for Smoothing Level}
As noted in our taxonomy of smoothing techniques, each method provides 1 or more input parameters for adjusting the level of smoothing. However, the input parameters for each technique have little to no direct relationship to any other technique. For example, \methodGau smoothing with $\sigma=5$ samples has no analog to \methodUni subsampling $50\%$ of points, even though they may produce similar results. This makes the analytical comparison of techniques difficult. 

Recently, approximate entropy ($ApEx$) was shown to be a high quality proxy for the visual complexity of line charts~\cite{2018entropy}. Generally speaking, approximate entropy is a measure that quantifies predictability of fluctuations in the data. In our case, we see visual complexity and smoothing level as synonymous---therefore, $ApEx$ is used as a baseline for our analysis. In other words, if the smoothed outputs of 2~different techniques have the same $ApEx$, we consider their smoothing level identical, e.g.,  
in \autoref{fig:teaser}, all methods have similar $ApEx$ values. See~\cite{2018entropy} for a description of how to compute $ApEx$.

\begin{figure}[!b]
    \centering

    \begin{minipage}[t]{0.3\linewidth}
        \includegraphics[height=2.0cm]{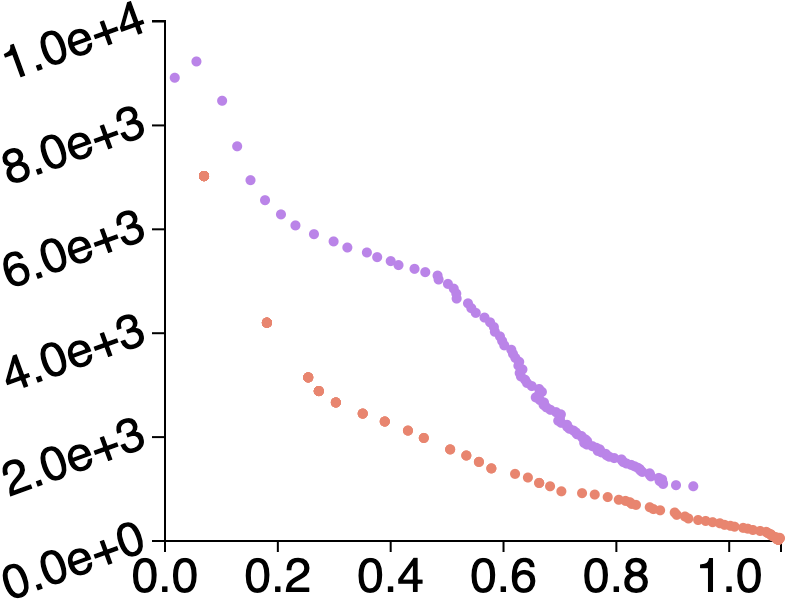}
    \end{minipage}
    \begin{minipage}[t]{0pt}
        \vspace{-60pt}
        \begin{minipage}[t]{15pt}
            \hspace{-25pt}
            \subfigure[\label{fig:entropy_plot:points}]{\hspace{15pt}}
        \end{minipage}
    \end{minipage}
    \hspace{5pt}
    \begin{minipage}[t]{0.25\linewidth}
        \includegraphics[trim= 133pt 0 0 0, clip, height=2.0cm]{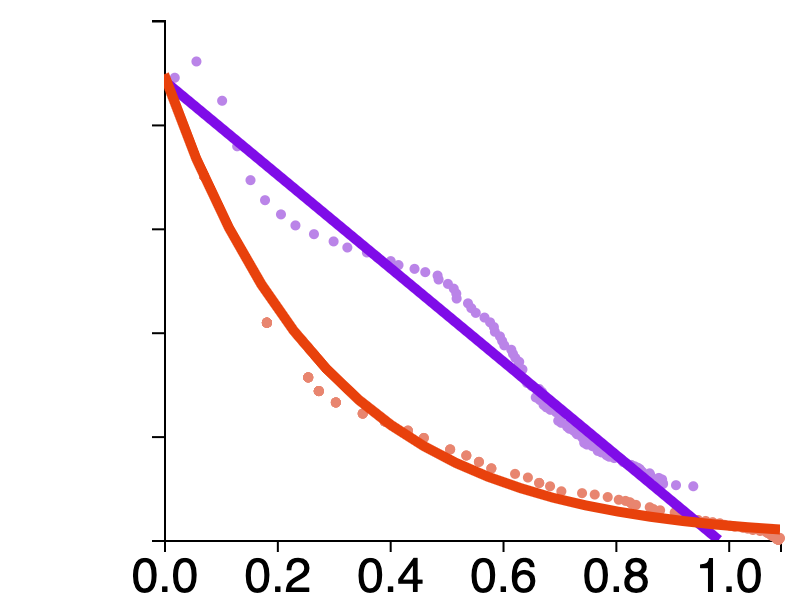}
    \end{minipage}
    \begin{minipage}[t]{0pt}
        \vspace{-60pt}
        \begin{minipage}[t]{15pt}
            \hspace{-25pt}
            \subfigure[\label{fig:entropy_plot:regression}]{\hspace{15pt}}
        \end{minipage}
    \end{minipage}
    \hspace{5pt}
    \begin{minipage}[t]{0.25\linewidth}
        \includegraphics[trim= 133pt 0 0 0, clip, height=2.0cm]{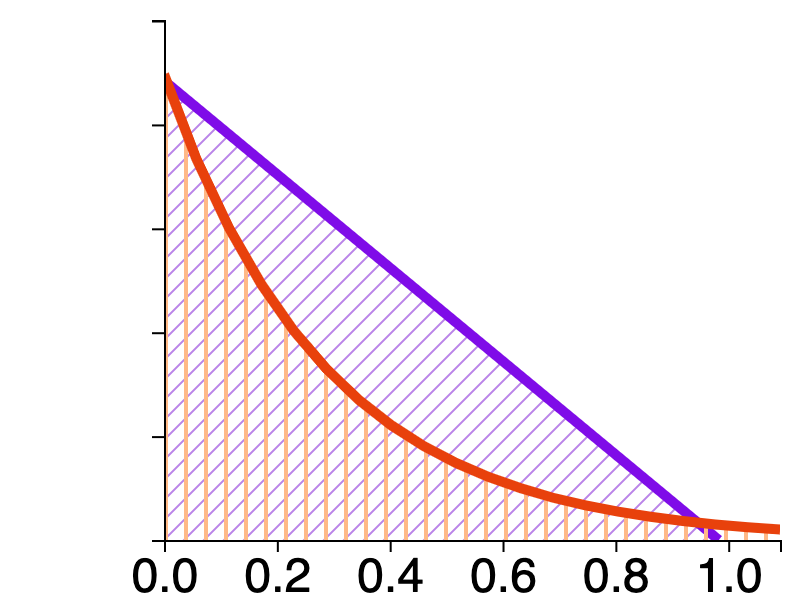}
    \end{minipage}
    \begin{minipage}[t]{0pt}
        \vspace{-60pt}
        \begin{minipage}[t]{15pt}
            \hspace{-25pt}
            \subfigure[\label{fig:entropy_plot:integration}]{\hspace{15pt}}
        \end{minipage}
    \end{minipage}

    \caption{Entropy plot for the EEG Channel 10 (500 samples) dataset with \methodTDA (in red) and \methodCheb (in purple).  The $L^1$-norm is plotted vertically, and the $ApEx$ horizontally. (a)~The methods are sampled at 100 levels to capture a range of $ApEx$ values. (b)~Linear and logarithmic regression are used to obtain a best-fit model. (c)~Mathematical integration is used to capture the area under the curve. Methods with smaller areas produce less error, thus performing better. In this case, \methodTDA is $\sim2500$, while \methodCheb is $\sim4350$, making \methodTDA the more effective method.}
    \label{fig:entropy_plot}
\end{figure}

\begin{figure*}[!t]
    \centering
    \includegraphics[width=\linewidth]{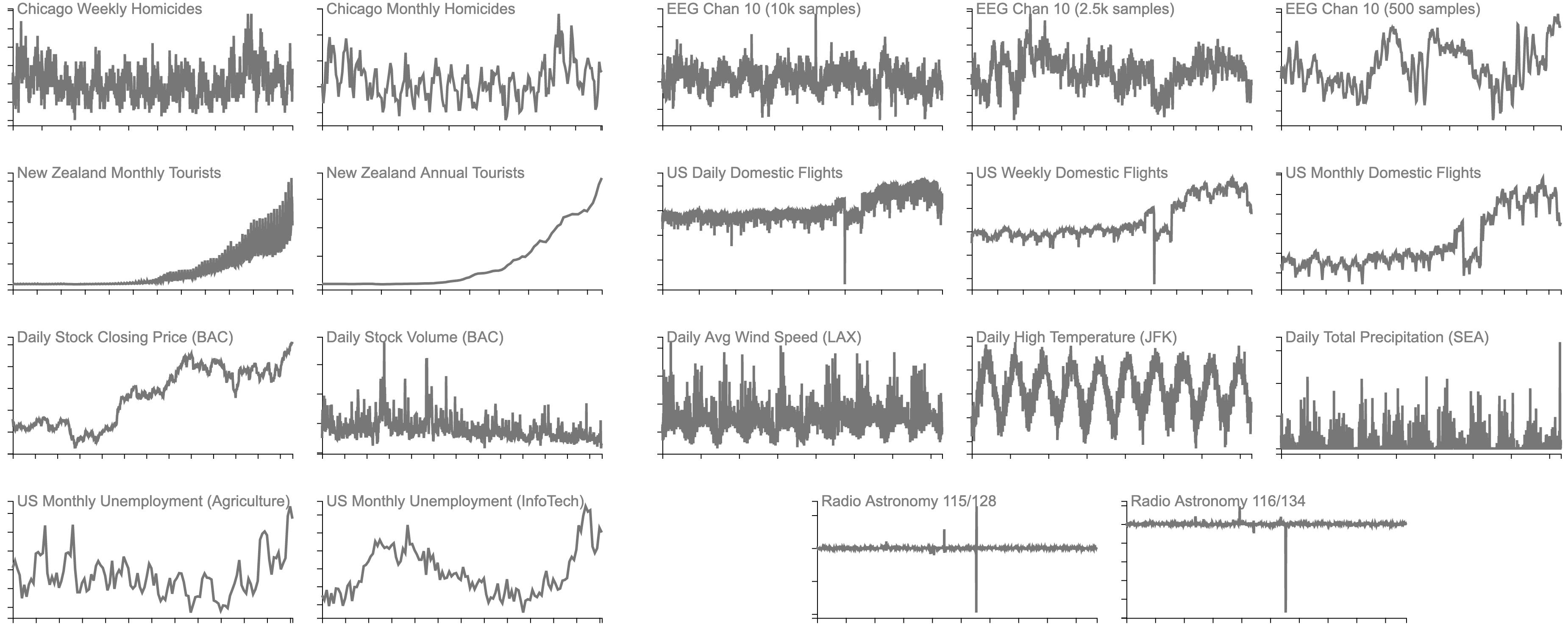}
    \begin{minipage}{0pt}
        \begin{minipage}{505pt}
            \vspace{-174pt}
            \hspace{-250pt}
            \subfigure[Weekly and Monthly Homicide Rates for Chicago\label{fig:datasets:chi}]{\hspace{185pt}}
            \hspace{20pt}
            \subfigure[Single channel of EEG Data with 10k, 2.5k, and 500 samples\label{fig:datasets:eeg}]{\hspace{285pt}}
            
            \vspace{31pt}
            \hspace{-250pt}
            \subfigure[Monthly and Annual Tourists in New Zealand\label{fig:datasets:nz}]{\hspace{185pt}}
            \hspace{20pt}
            \subfigure[Daily, Weekly, and Monthly US Domestic Flights\label{fig:datasets:flights}]{\hspace{285pt}}
            
            \vspace{31pt}
            \hspace{-250pt}
            \subfigure[Daily Stock Closing Price and Trading Volume\label{fig:datasets:stock}]{\hspace{185pt}}
            \hspace{20pt}
            \subfigure[Daily Avg Wind Speed, High Temperature, and Total Precipitation at US Airports\label{fig:datasets:climate}]{\hspace{285pt}}
            
            \vspace{31pt}
            \hspace{-250pt}
            \subfigure[US Monthly Unemployment by Sector\label{fig:datasets:unemp}]{\hspace{185pt}}
            \hspace{20pt}
            \subfigure[Radio Astronomy Signals\label{fig:datasets:astro}]{\hspace{285pt}}
            \hspace{100pt}
        \end{minipage}
    \end{minipage}

    \vspace{-10pt}    
    \caption{Examples of datasets used to analyze smoothing techniques.}
    \label{fig:datasets}
\end{figure*}

\subsubsection{Ranking the Effectiveness of Smoothing}

To select the most effective technique for a given metric, we want to focus on those that have the smallest value. 

\paragraph{Ranking a Single Smoothing Level}
When smoothing results have equivalent $ApEx$, ranking their effectiveness is fairly trivial. For a given metric, the techniques are simply ordered from lowest (best) to highest (worst). For example, in \autoref{fig:teaser:ranks}, the $L^1$-norm, $\ell_1$, in the first column, shows that \methodTDA has the lowest error, followed by \methodGau and \methodSG. The rankings can be computed for all metrics, and the smoothing techniques evaluated for all tasks. For example, in \autoref{fig:teaser:ranks}, \methodSG is in the top 3 for all metrics/tasks, making it a reasonable choice to represent the input data.

\begin{figure}[!b]
    \centering
    \hfill
    \includegraphics[width=0.825\linewidth]{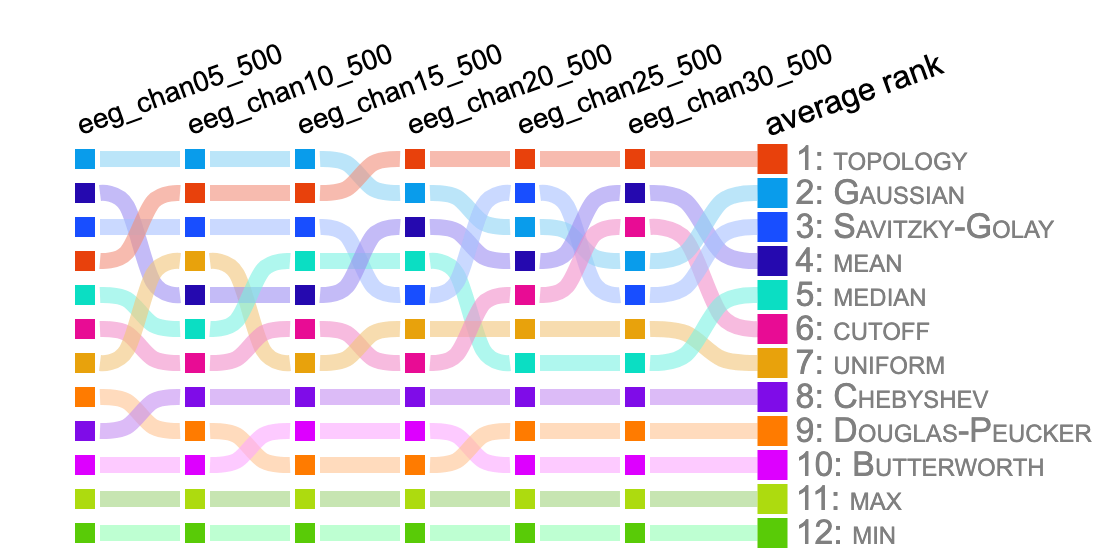}
    \hspace{15pt}
    
    \caption{Rank plot of the $L^1$-norm for all EEG (500 samples) datasets. Each column ranks the techniques, from best (top) to worst (bottom), on a different dataset, with the average rank in the final column. The result shows that for this data, on average, \methodTDA, \methodGau, and \methodSG are the most effective techniques, respectively.}
    \label{fig:dataset_summary}
\end{figure}

\paragraph{Ranking All Smoothing Levels}
Summarizing the performance of different smoothing techniques across all smoothing levels requires additional analysis. We calculate 100 different smoothing levels, across a range of entropy values. For each metric, we create an entropy plot, which is a scatterplot of the metric value against a range of $ApEx$ values. In \autoref{fig:entropy_plot:points}, the $L^1$-norm is plotted vertically, against the $ApEx$ horizontally for \methodTDA, in red, and \methodCheb, in purple. 

Next, both linear and logarithmic regression are performed using iterative reweighted least-squares (IRLS)\footnote{Using IRLS helps to minimize the impact of outliers.}~\cite{holland1977robust}. The model, linear or logarithmic, with the larger $R^2$ value is selected as a proxy for the efficacy of the technique. For \autoref{fig:entropy_plot:regression}, \methodTDA is best modeled logarithmically, and \methodCheb is best modeled linearly. 

Since the goal is again to minimize the error induced in the data, the total area under the regression is computed (i.e., mathematical integration), and the methods are ranked smallest to largest area. In \autoref{fig:entropy_plot:integration}, the total area is $\sim2500$ for \methodTDA and $\sim4350$ for \methodCheb, making \methodTDA more effective than \methodCheb. 

The resulting ranks are placed into a rank plot, as seen in \autoref{fig:dataset_summary}. In this plot, the $L^1$-norm is ranked across multiple datasets. Each dataset receives a column, and the smoothing methods are ranked from best (top) to worst (bottom). The tracks are added to improve readability.

\paragraph{Ranking Across Multiple Datasets}
To summarize the overall efficacy of techniques across multiple datasets, an average rank is calculated. The average rank simply takes the sum of the rank across all datasets and orders them from best (top) to worst (bottom). In \autoref{fig:dataset_summary}, the average rank is the final column. The average rank of \methodTDA is $\frac{4+2+2+1+1+1}{6}=1.8$, while the \methodGau is $2.0$, \methodSG is $3.5$, etc.

%% file: sec-results.tex
\section{Results}
\label{sec:eval}

The source code and data for our framework and evaluation is available at {\small\textless\textcolor{blue}{\url{https://github.com/USFDataVisualization/LineSmooth}}\textgreater}, and an interactive version of our evaluation is at {\small\textless\textcolor{blue}{\url{https://usfdatavisualization.github.io/LineSmoothDemo}}\textgreater}.

\subsection{Data Sources}

We evaluate 80 datasets in 13 categories from 8 data sources (see \autoref{fig:datasets}). Data were selected that contained typical qualities, such as long-term trends (e.g., stock prices), cyclical behaviors (e.g., daily temperatures), or important spikes (e.g., a stock market crash).
\begin{itemize}

    \item \underline{Chicago Homicide Rates} (\textit{chi\_homicide}) data (see \autoref{fig:datasets:chi}) contains weekly (969 samples) and monthly (222 samples) counts of the number of homicides in the city from January 2001 through July 2019. Data is provided by the City of Chicago~\cite{chicago_crime}.

    \item \underline{EEG} (\textit{eeg\_500}, \textit{eeg\_2500}, and \textit{eeg\_10000}) data (see \autoref{fig:datasets:eeg}) contains windows of 3 different lengths (500, 2500, and 10000 samples) from 6 (of 32 total) channels from a single subject undergoing a visual attention task and was acquired from the EEG/ERP Public Archive~\cite{eeg_database}.

    \item \underline{New Zealand Tourist} (\textit{nz\_tourist}) data (see \autoref{fig:datasets:nz}) contains the monthly (1165 samples) and annual (96 samples) number of tourists visiting the country from April 1921 through April 2018. Data collected from Trading Economics~\cite{new_zealand}.

    \item \underline{US Domestic Flights} (\textit{flights}) data (see \autoref{fig:datasets:flights}) contains the number of daily, weekly, and monthly (7671, 1095, and 252 samples, respectively) number of US flights from January 1, 1988 through December 31, 2008. Data collected from Observable~\cite{d3_zoomable}.

    \item \underline{Stock Price} (\textit{stock\_price}) and \underline{Stock Volume} (\textit{stock\_volume}) data (see \autoref{fig:datasets:stock}) contains daily closing values and trading volumes, respectively, for 9 companies (Apple, Amazon, Bank of America, Google, Intel, JP Morgan, Microsoft, Toyota, and Tesla) over a 5 year period, January 2015 through December 2019 (1257 samples each), collected from Yahoo Finance~\cite{yahoo_finance}.
    
    \item \uline{Average Wind Speed} (\textit{climate\_awnd}), \uline{High Temperature} (\textit{climate\_tmax}), and \uline{Total Precipitation} (\textit{climate\_prcp}) data (see \autoref{fig:datasets:climate}) contains 10 years of daily weather values (3651 samples each) from 6 US Airports (Atlanta, New York JFK, Los Angeles, Chicago O'Hare, Seattle-Tacoma, and Salt Lake City), collected from NOAA Climate Data Service~\cite{noaa}.

    \item \underline{US Unemployment} (\textit{unemployment}) data (see \autoref{fig:datasets:unemp}) are the monthly number of unemployed individuals in 14 economic sectors (e.g., agriculture, finance, health, etc.) from January 2000 through February 2010 (122 samples each). The data were collected from the US Bureau of Labor Statistics~\cite{us_bls}.

    \item \underline{Radio Astronomy} (\textit{astro}) data (see \autoref{fig:datasets:astro}) contains 5 spectral ``lines'' (1947 samples each) that measure the frequency and amplitude of radio waves emitted by extraterrestrial matter (i.e., gas and dust), collected from the ALMA Science Archive~\cite{alma}.
    
\end{itemize}

\subsection{Evaluation By Task}

We evaluate 12 smoothing methods from \autoref{table:filters} with our framework (see \autoref{fig:teaser}).
Tasks that use the same metrics are combined to reduce space. We produce rank plots summarizing all datasets for the metrics related to those tasks. Each column is the average rank for a given data category, with the average rank across all 80 datasets (i.e., total performance) in the final column. To reduce the clutter, we only show the tracks for smoothing methods with the top 4 overall performance.

\subsubsection{Retrieve Value / Determine Range}

\begin{figure}[!hbt]
    \centering
    \begin{minipage}[t]{0.05\linewidth}
        \rotatebox{90}{
		    \begin{minipage}{3cm}\subfigure[$L^1$-Norm  ($\ell_1$)]{\hspace{2.5cm}}\end{minipage}
		}
	\end{minipage}
	\hfill
	\begin{minipage}[t]{0.94\linewidth}
		\includegraphics[width=\linewidth]{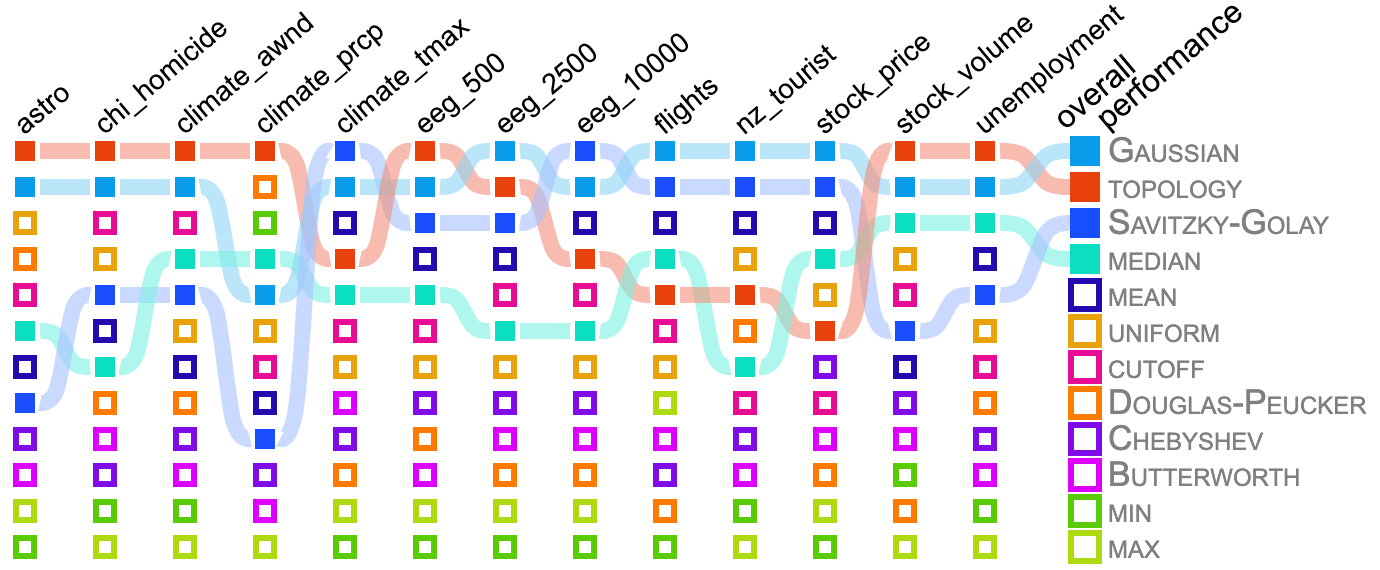}
	\end{minipage}
	
	\vspace{-10pt}
    \begin{minipage}[t]{0.05\linewidth}
        \rotatebox{90}{
		    \begin{minipage}{3cm}\subfigure[$L^\infty$-Norm ($\ell_\infty$)]{\hspace{2.5cm}}\end{minipage}
		}
	\end{minipage}
	\hfill
	\begin{minipage}[t]{0.94\linewidth}	
        \includegraphics[trim=0 0 0 134pt, clip, width=\linewidth]{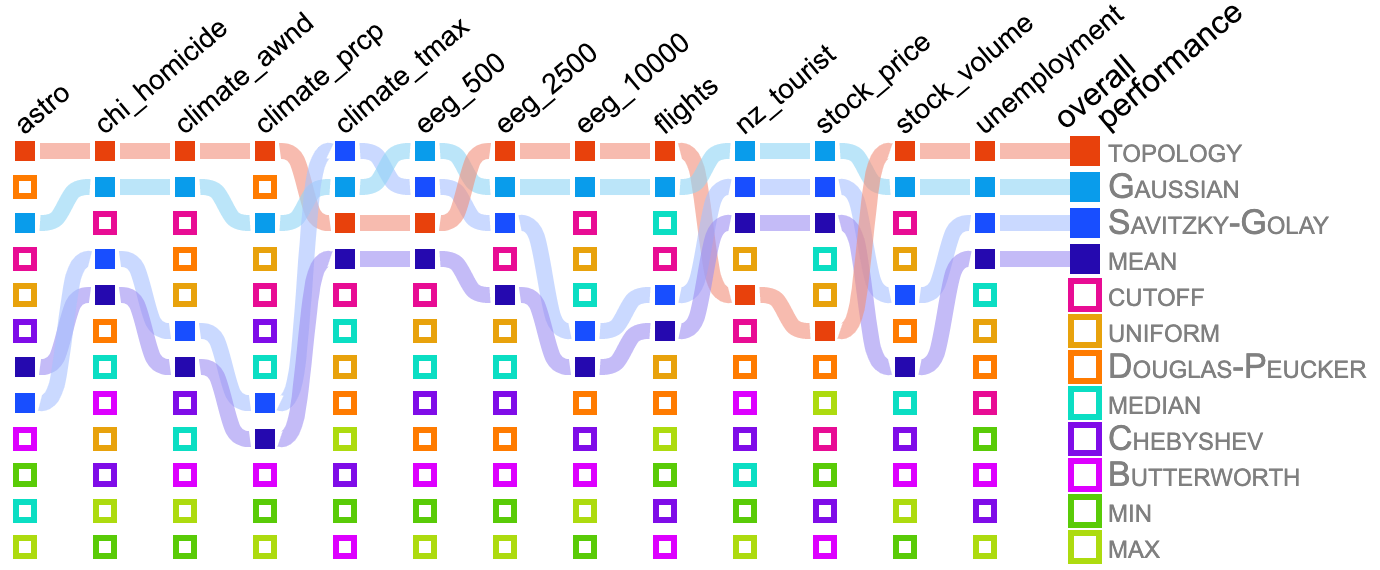}
	\end{minipage}
        
    \caption{Average ranking for the \textbf{Retrieve Value} and \textbf{Determine Range} tasks across all datasets using the (a)~$L^1$-norm and (b)~$L^\infty$-norm. Both metrics show that \methodTDA and \methodGau were primarily the best methods to use, while \methodSG occasionally performs well.}
    \label{fig:results:rv}
\end{figure}

The results for \textbf{Retrieve Value} and \textbf{Determine Range} tasks, in \autoref{fig:results:rv}, show that 3 smoothing techniques---\methodTDA, \methodGau, and \methodSG---produced the best results. Still among those datasets, there is a distinction between results depending upon the dataset category. For example, \methodTDA excelled at data that are predominated by ``spiky'' features, e.g., \textit{climate\_prcp} and \textit{stock\_volume}. On the other hand, \methodGau and \methodSG performed better on data with long-term trends, e.g., \textit{stock\_price}, and cyclical behaviors, e.g., \textit{climate\_tmax}.
Among the worst performing techniques were all rank-based approaches, all frequency domain-based approaches, and the \methodUni and \methodDP subsampling approaches.

\subsubsection{Compute Derived Value}

\begin{figure}[!ht]
    \centering
    \hfill
    \includegraphics[width=0.94\linewidth]{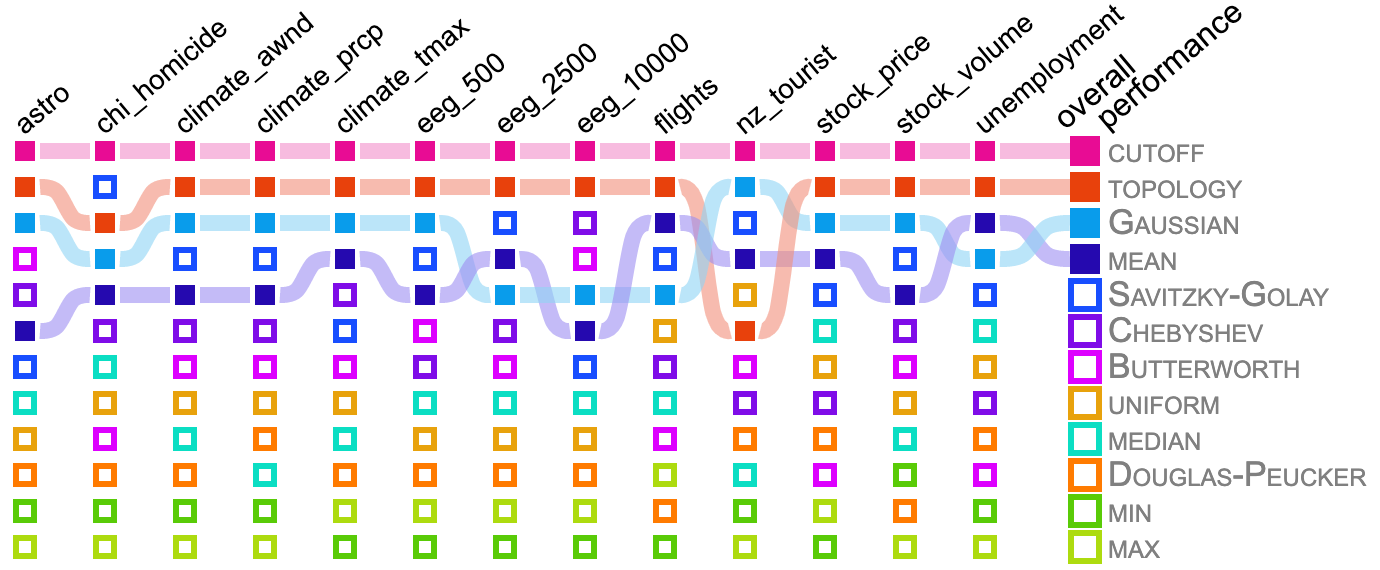}
    \caption{Average ranking for the \textbf{Compute Derived Value} task across all datasets using the area preservation metric, $\delta a$. The results show that \methodCut is clearly the top method, followed by \methodTDA. \methodGau, \methodMean, and \methodSG occasionally perform well.}
    \label{fig:results:cdv}
\end{figure}

The results for the \textbf{Computing Derived Value} task, in \autoref{fig:results:cdv}, show that the \methodCut filter clearly outperforms all other techniques. Examining the entropy plots (not shown), it can be observed that \methodTDA performs similarly well on most of the data sets. Finally, the convolutional methods, \methodGau, \methodMean, and \methodSG, occasionally performed well.
Among the worst performing techniques are all rank-based techniques, frequency domain-based techniques, excluding \methodCut, and subsampling techniques, excluding \methodTDA.

\subsubsection{Find Extrema / Find Anomalies}

\begin{figure}[!hb]
    \centering
    \begin{minipage}[t]{0.05\linewidth}
        \rotatebox{90}{
		    \begin{minipage}{3cm}\subfigure[Wasserstein ($W_1$)\label{fig:results:fe:avg}]{\hspace{2.5cm}}\end{minipage}
		}
	\end{minipage}
	\hfill
	\begin{minipage}[t]{0.94\linewidth}
		\includegraphics[width=\linewidth]{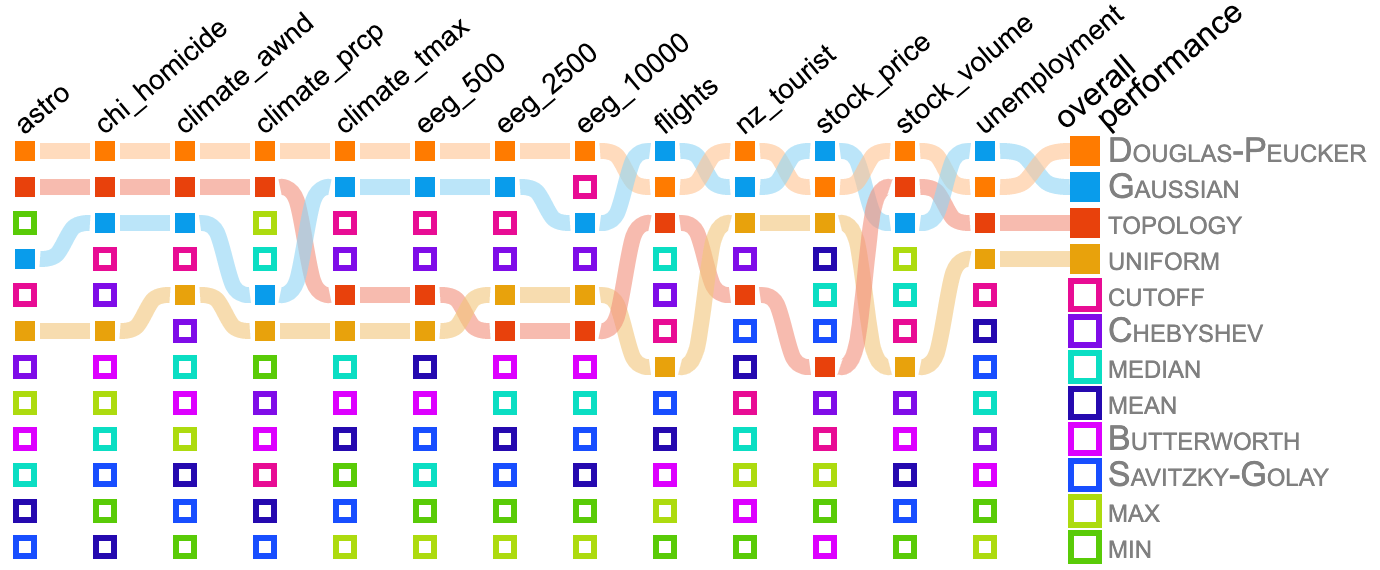}
	\end{minipage}
	
	\vspace{-10pt}
    \begin{minipage}[t]{0.05\linewidth}
        \rotatebox{90}{
		    \begin{minipage}{3cm}\subfigure[Bottleneck ($W_\infty$)\label{fig:results:fe:worst}]{\hspace{2.5cm}}\end{minipage}
		}
	\end{minipage}
	\hfill
	\begin{minipage}[t]{0.94\linewidth}	
        \includegraphics[trim=0 0 0 134pt, clip, width=\linewidth]{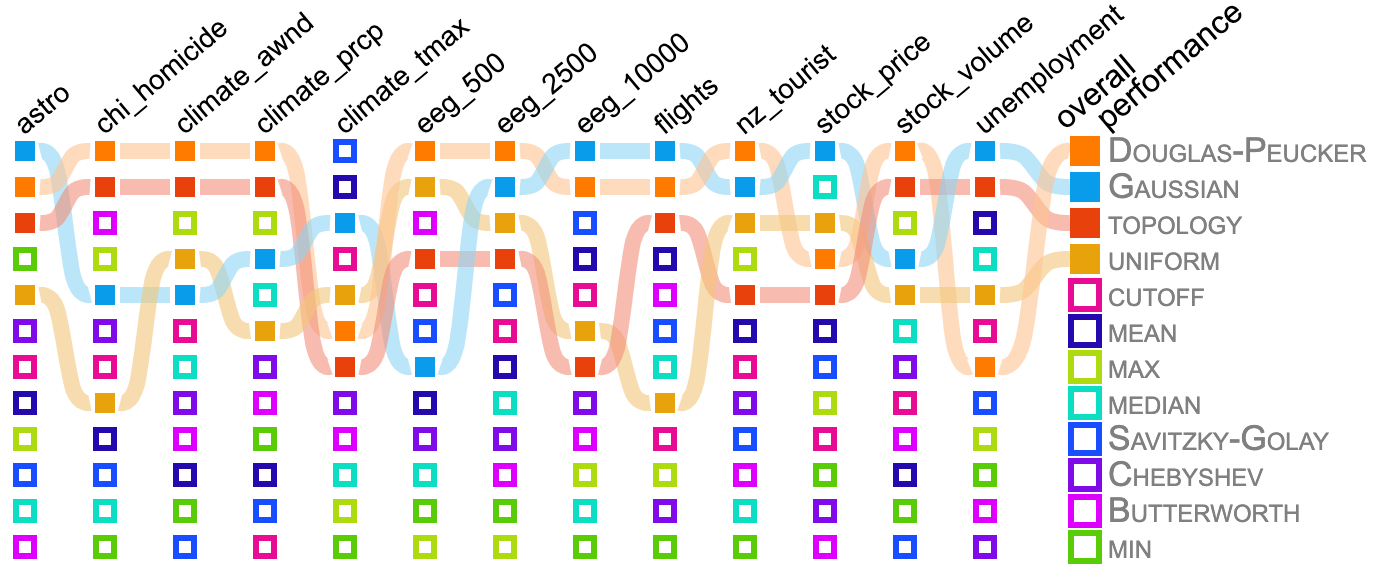}
	\end{minipage}

    \caption{Average ranking for the \textbf{Find Extrema} and \textbf{Find Anomalies} tasks across all datasets using the (a) average case, Wasserstein distance, and (b) worst case, Bottleneck distance. Both metrics show \methodDP performing best, followed by \methodGau, \methodTDA, and \methodUni, in order.
    }
    \label{fig:results:fe}
\end{figure}

The results for \textbf{Find Extrema} and \textbf{Find Anomalies} tasks, in \autoref{fig:results:fe}, show that \methodDP produced the best results, with \methodGau, \methodTDA, and, interestingly, \methodUni occasionally performing well. Since the subsampling techniques, including \methodUni, use a subset of the original data, it is safe to assume that for some levels of smoothing, peaks will be included in the output.
Among the techniques that performed poorly were once again rank-based, frequency domain-based, and convolutional techniques, excluding \methodGau.

\subsubsection{Characterize Distribution / Cluster: Trends}

\begin{figure}[!ht]
    \centering
    \hfill
    \includegraphics[width=0.94\linewidth]{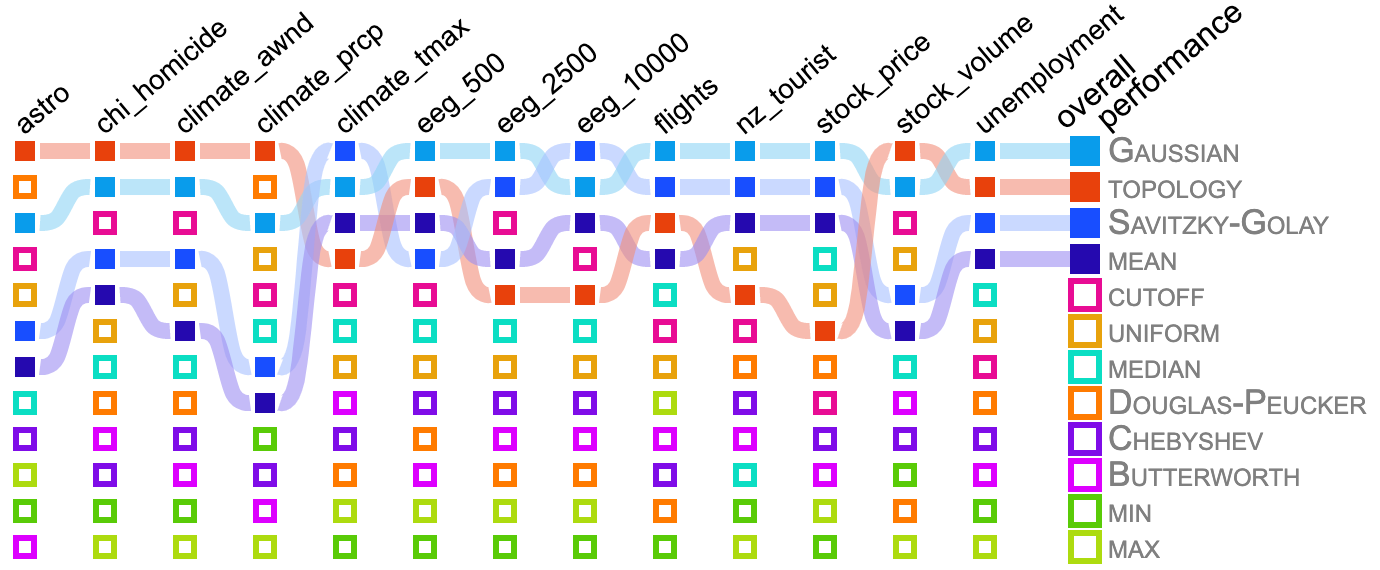}
    \caption{Average ranking for the \textbf{Characterize Distribution} and \textbf{Cluster: Trends} tasks across all datasets using the frequency preservation metric, $\Fgroup$. The results show that depending upon the type of data, \methodGau, \methodTDA, and \methodSG methods produce the best results.}
    \label{fig:results:cd}
\end{figure}

The results for the \textbf{Characterize Distribution} and \textbf{Cluster: Trends} task, in \autoref{fig:results:cd}, show that depending upon the datasets, \methodGau, \methodTDA, or \methodSG produce the best results, with \methodTDA working best on ``spiky'' datasets and convolutional techniques working better on those datasets with long-term trends and cyclical behaviors.
Among the worst performing techniques are again all rank-based techniques, frequency domain-based techniques, and subsampling, excluding \methodTDA.

\subsubsection{Sort / Cluster: Points}

\begin{figure}[!t]
    \centering
    
    \centering
    \begin{minipage}[t]{0.05\linewidth}
        \rotatebox{90}{
		    \begin{minipage}{3cm}\subfigure[Pearson ($\rho$)\label{fig:results:s:pcc}]{\hspace{2.5cm}}\end{minipage}
		}
	\end{minipage}
	\hfill
	\begin{minipage}[t]{0.94\linewidth}
		\includegraphics[width=\linewidth]{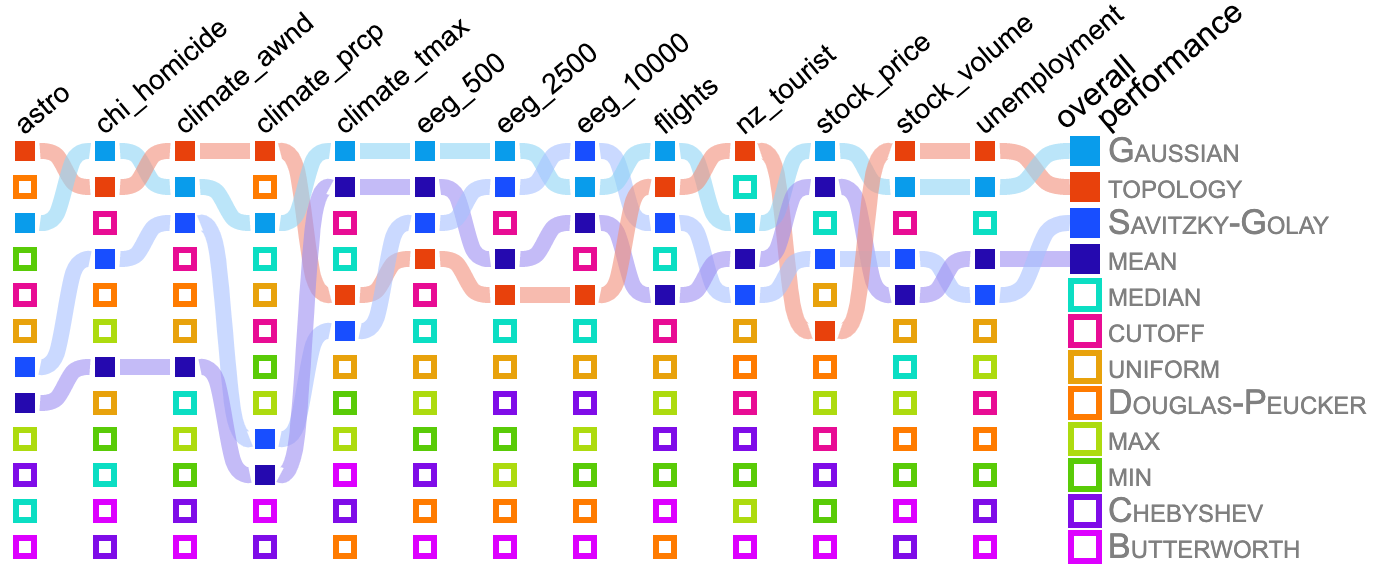}
	\end{minipage}
	
	\vspace{-10pt}
    \begin{minipage}[t]{0.05\linewidth}
        \rotatebox{90}{
		    \begin{minipage}{3cm}\subfigure[Spearman ($r_s$)\label{fig:results:s:src}]{\hspace{2.5cm}}\end{minipage}
		}
	\end{minipage}
	\hfill
	\begin{minipage}[t]{0.94\linewidth}	
        \includegraphics[trim=0 0 0 134pt, clip, width=\linewidth]{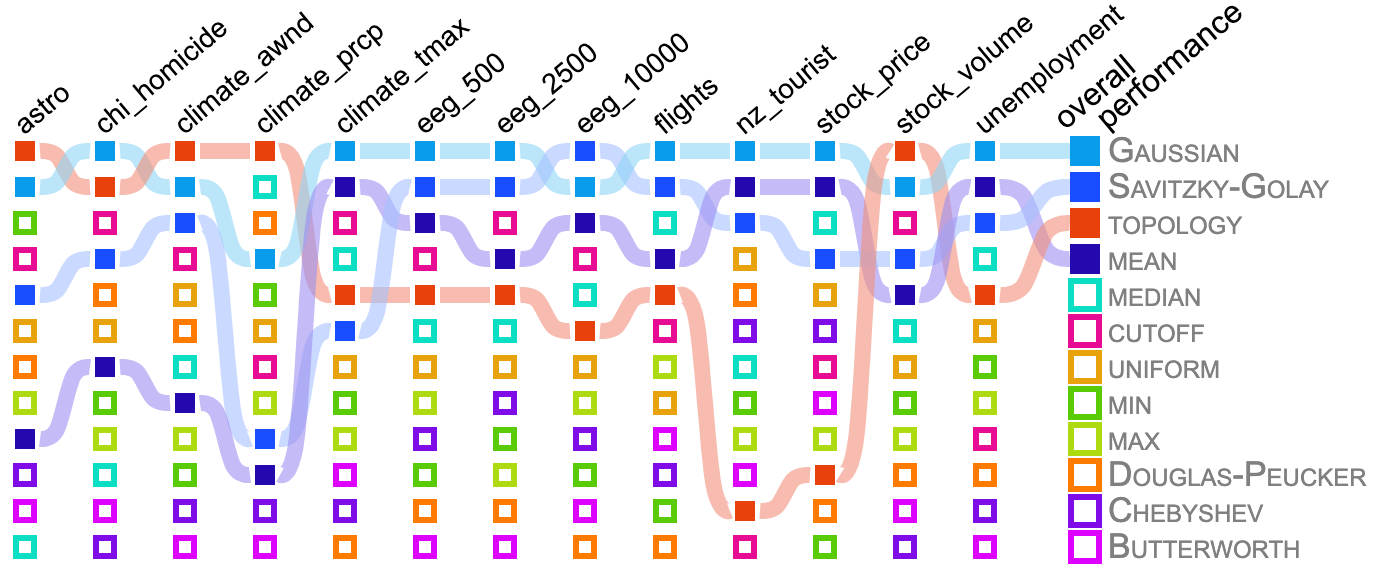}
	\end{minipage}

    \caption{Average ranking for the \textbf{Sort} and \textbf{Cluster: Points} tasks across all datasets using the (a)~Pearson Correlation Coefficient, $\rho$, and (b)~Spearman Rank Correlation, $r_s$. Both metrics show \methodGau performing best, followed by \methodTDA and \methodSG.}
    \label{fig:results:s}
\end{figure}

The results for the \textbf{Sort} and \textbf{Cluster: Points} tasks, in \autoref{fig:results:s}, show that \methodGau was the best performer, followed by \methodTDA and \methodSG, depending upon whether the Pearson metric, in \autoref{fig:results:s:pcc}, or the Spearman metric, in \autoref{fig:results:s:src}, is used. Once again, the \methodTDA method appears to work best on ``spiky'' datasets, while the convolutional methods worked better on data with long-term trends or cyclical behaviors.
The worst performing techniques for these tasks are largely the same as for other tasks.

%% file: sec-discussion.tex
\section{Discussion: Smoothing Recommendations}

Given our evaluation in \autoref{sec:eval}, we summarize the efficacy of the techniques tested in \autoref{tbl:grades}. For these grades, we measure the frequency of a method being ranked in the top 3 for a given task across each of the 80 datasets. Grades are assigned using that frequency: A: $>75\%$; B: $50\%-75\%$; C: $25\%-50\%$; D: $5\%-25\%$. In other words, a method scoring a grade of A ranks 1st, 2nd, or 3rd in at least $75\%$ of datasets.

\paragraph{General Recommendation} If designing a visualization without particular concern for the data or visual analytics task, \methodGau and \methodTDA performed well in all categories. The main difference between the 2 is that while \methodGau had more A's, \methodTDA has no score lower than B. The heart of which to pick really lies in the type of data being visualized. As we pointed out in our evaluation, \methodTDA tended to do better on data with ``spiky'' features, while \methodGau did better with cyclical behaviors or long-term trends.

\newcommand{\SA}{\multicolumn{2}{c|}{\fontfamily{\sfdefault}\selectfont\textbf{A}}}
\newcommand{\SB}{\multicolumn{2}{c|}{\fontfamily{\sfdefault}\selectfont\textbf{B}}}
\newcommand{\SC}{\multicolumn{2}{c|}{\fontfamily{\sfdefault}\selectfont C}}
\newcommand{\SD}{\multicolumn{2}{c|}{\fontfamily{\sfdefault}\selectfont D}}
\newcommand{\SX}{\multicolumn{2}{c|}{--}}

\newcommand{\theaderC}[1]{\multicolumn{2}{c}{
    \begin{minipage}[b]{20pt}
        \hspace{-90pt}
        {\begin{minipage}[b]{0pt}\rotatebox[origin=l]{-35}{{\begin{minipage}[b]{115pt}\begin{flushright}#1\end{flushright}\end{minipage}}}\end{minipage}}
        \end{minipage}}}

\renewcommand*{\arraystretch}{1.5}
\begin{table}[!t]
    \centering
    \caption{Grades for the efficacy of different smoothing methods are calculated by the frequency of being ranked in the top 3 for all 80 datasets. A: $>75\%$; B: $50\%-75\%$; C: $25\%-50\%$; D: $5\%-25\%$}
    \label{tbl:grades}
    
    \vspace{-15pt}
    {\resizebox{1.0\linewidth}{!}{%
\begin{tabular}{r|cc|cc|cc|cc|cc|cc|cc|cc|cc|}
   \multicolumn{1}{c}{}
 & \theaderC{Retrieve Value} 
 & \theaderC{Determine Range} 
 & \theaderC{Compute Derived Value}  
 & \theaderC{Find Extrema} 
 & \theaderC{Find Anomalies} 
 & \theaderC{Characterize Distribution} 
 & \theaderC{Sort} 
 & \theaderC{Cluster: Trends}
 & \theaderC{Cluster: Points}\\ 
 
\methodMed  & \SD & \SD & \SX & \SD & \SD & \SD & \SC & \SD & \SC \\
\hline
\methodMin  & \SX & \SX & \SX & \SX & \SX & \SX & \SX & \SX & \SX \\
\hline
\methodMax  & \SX & \SX & \SX & \SD & \SD & \SX & \SX & \SX & \SX \\ 
\hline
\methodGau  & \SA & \SA & \SC & \SB & \SB & \SA & \SA & \SA & \SA \\
\hline
\methodSG   & \SC & \SC & \SD & \SD & \SD & \SB & \SC & \SB & \SC \\
\hline
\methodMean & \SC & \SC & \SD & \SD & \SD & \SC & \SC & \SC & \SC \\
\hline
\methodCut  & \SD & \SD & \SA & \SD & \SD & \SC & \SD & \SC & \SD \\
\hline
\methodButt & \SX & \SX & \SD & \SD & \SD & \SX & \SX & \SX & \SX \\
\hline
\methodCheb & \SX & \SX & \SD & \SX & \SX & \SX & \SX & \SX & \SX \\
\hline
\methodUni  & \SX & \SX & \SX & \SD & \SD & \SX & \SX & \SX & \SX \\
\hline
\methodDP   & \SD & \SD & \SX & \SA & \SA & \SD & \SD & \SD & \SD \\ 
\hline
\methodTDA  & \SB & \SB & \SA & \SB & \SB & \SB & \SB & \SB & \SB \\ 
\hline

\end{tabular}
}}
\end{table}

\paragraph{Task Specific Recommendations} If the visual analytics tasks are known ahead of time, a more nuanced decision can be made about what method to use. While \methodGau and \methodTDA did well on most tasks, \methodCut and \methodDP performed the best on \textbf{Compute Derived Value} and \textbf{Find Extrema/Anomalies}, respectively.

\paragraph{Data Specific Recommendations} If the data are available ahead of time, analyzing them with our framework to select the best smoothing method is recommended. In addition to \methodGau and \methodTDA, several methods generate better results in limited situations, particularly if the tasks are known as well. These methods include \methodMed, \methodSG, \methodMean, \methodCut, and \methodDP.

\paragraph{Methods to Largely Avoid} Several methods performed poorly across the board. These include \methodMin, \methodMax, \methodButt, \methodCheb, and \methodUni subsampling. These methods rarely performed in the top 3. They should only be used when there is a very specific reason to do so, which should not be a problem as most of these techniques are rarely used anyways. However, this finding is particularly relevant for \methodUni subsampling, as it is essentially the default methodology used for data reduction~\cite{hansen2011visualization}.

\subsection{Conclusions}

In conclusion, we have presented and demonstrated a framework for evaluating line chart smoothing in the context of the visual analytics tasks being performed. There remain several study limitations and future works.

\paragraph{Perceptual Effects and User-based Validation} Our study considers only the effects of data modification in the evaluation of smoothing effectiveness. There may be additional perceptual effects that make the results of some techniques better or worse than others. 
We considered performing a user study to validate our framework further. However, it became quickly apparent that the scale of such a study would be impractical. As an example, testing 12 smoothing techniques, across 8 tasks, 80 datasets, and 20 different smoothing levels would require in excess of 150k experimental stimuli.

\paragraph{Feature Types and Representing Lost Information} Throughout our analysis, we discussed ``spiky'', cyclical, and long-term trends in data. These categories are ill-defined, and a broader study of feature types that appear in line charts would be valuable to the community.
Furthermore, smoothing removes information from the representation of the line chart, introducing uncertainty. One additional direction of future work would be to use this framework to model and represent the uncertainty while considering the context of the visual analytics task.

\paragraph{There's No Accounting for Taste} Aesthetics play an important role in visualization design. Without a good aesthetic, users are less likely to remember what they see~\cite{borkin2013makes}. Although \methodTDA and \methodGau were largely the most effective techniques, their aesthetics are quite different, ``spiky'' for \methodTDA and smooth for \methodGau. Our framework completely ignores aesthetic in its recommendation, in part because aesthetic is both art and science, thus difficult to model mathematically or algorithmically.

%% file: main.bbl
\begin{thebibliography}{10}

\bibitem{2016adnan}
M.~Adnan, M.~Just, and L.~Baillie.
\newblock Investigating time series visualisations to improve the user
  experience.
\newblock In {\em ACM SIGCHI Conference on Human Factors in Computing Systems},
  pp. 5444--5455, 2016. doi: {{%
10\hspace{.1pt}\discretionary{.}{%
}{.}\hspace{.4pt}1145\discretionary{/}{%
}{/}2858036\hspace{.1pt}\discretionary{.}{%
}{.}\hspace{.4pt}2858300}}


\bibitem{albers2014task}
D.~Albers, M.~Correll, and M.~Gleicher.
\newblock Task-driven evaluation of aggregation in time series visualization.
\newblock In {\em ACM SIGCHI Conference on Human Factors in Computing Systems},
  pp. 551--560, 2014. doi: {{%
10\hspace{.1pt}\discretionary{.}{%
}{.}\hspace{.4pt}1145\discretionary{/}{%
}{/}2556288\hspace{.1pt}\discretionary{.}{%
}{.}\hspace{.4pt}2557200}}


\bibitem{alma}
{Alma Science Archive}.
\newblock \url{http://almascience.nrao.edu/aq/}, Apr 2020.

\bibitem{amar2005low}
R.~Amar, J.~Eagan, and J.~Stasko.
\newblock Low-level components of analytic activity in information
  visualization.
\newblock In {\em IEEE Symposium on Information Visualization (InfoVis)}, pp.
  111--117, 2005. doi: {{%
10\hspace{.1pt}\discretionary{.}{%
}{.}\hspace{.4pt}1109\discretionary{/}{%
}{/}INFVIS\hspace{.1pt}\discretionary{.}{%
}{.}\hspace{.4pt}2005\hspace{.1pt}\discretionary{.}{%
}{.}\hspace{.4pt}1532136}}


\bibitem{2016arbesser}
C.~{Arbesser}, F.~{Spechtenhauser}, T.~{M{\"u}hlbacher}, and H.~{Piringer}.
\newblock Visplause: Visual data quality assessment of many time series using
  plausibility checks.
\newblock {\em IEEE Transactions on Visualization and Computer Graphics},
  23(1):641--650, 2017. doi: {{%
10\hspace{.1pt}\discretionary{.}{%
}{.}\hspace{.4pt}1109\discretionary{/}{%
}{/}TVCG\hspace{.1pt}\discretionary{.}{%
}{.}\hspace{.4pt}2016\hspace{.1pt}\discretionary{.}{%
}{.}\hspace{.4pt}2598592}}


\bibitem{arce2005nonlinear}
G.~R. Arce.
\newblock {\em Nonlinear signal processing: a statistical approach}.
\newblock John Wiley \& Sons, 2005. doi: {{%
10\hspace{.1pt}\discretionary{.}{%
}{.}\hspace{.4pt}1002\discretionary{/}{%
}{/}0471691852}}


\bibitem{best2007perception}
L.~A. Best, L.~D. Smith, and D.~A. Stubbs.
\newblock Perception of linear and nonlinear trends: using slope and curvature
  information to make trend discriminations.
\newblock {\em Perceptual and Motor Skills}, 2007. doi: {{%
10\hspace{.1pt}\discretionary{.}{%
}{.}\hspace{.4pt}2466\discretionary{/}{%
}{/}pms\hspace{.1pt}\discretionary{.}{%
}{.}\hspace{.4pt}104\hspace{.1pt}\discretionary{.}{%
}{.}\hspace{.4pt}3\hspace{.1pt}\discretionary{.}{%
}{.}\hspace{.4pt}707\discretionary{%
}{-}{-}721}}


\bibitem{borkin2013makes}
M.~A. Borkin, A.~A. Vo, Z.~Bylinskii, P.~Isola, S.~Sunkavalli, A.~Oliva, and
  H.~Pfister.
\newblock What makes a visualization memorable?
\newblock {\em IEEE Transactions on Visualization and Computer Graphics},
  19(12):2306--2315, 2013. doi: {{%
10\hspace{.1pt}\discretionary{.}{%
}{.}\hspace{.4pt}1109\discretionary{/}{%
}{/}TVCG\hspace{.1pt}\discretionary{.}{%
}{.}\hspace{.4pt}2013\hspace{.1pt}\discretionary{.}{%
}{.}\hspace{.4pt}234}}


\bibitem{d3_zoomable}
M.~Bostack.
\newblock {Observable: Zoomable Area Chart}.
\newblock \url{https://observablehq.com/@d3/zoomable-area-chart}, Apr 2020.

\bibitem{butterworth1930theory}
S.~Butterworth et~al.
\newblock On the theory of filter amplifiers.
\newblock {\em Wireless Engineer}, 7(6):536--541, 1930.

\bibitem{chicago_crime}
{Chicago Police Department}.
\newblock {Crimes - 2001 to present: City of Chicago: Data Portal}.
\newblock
  \url{https://data.cityofchicago.org/Public-Safety/Crimes-2001-to-present/ijzp-q8t2},
  Apr 2020.

\bibitem{chin1983quantitative}
R.~T. Chin and C.-L. Yeh.
\newblock Quantitative evaluation of some edge-preserving noise-smoothing
  techniques.
\newblock {\em Computer Vision, Graphics, and Image Processing}, 23(1):67--91,
  1983. doi: {{%
10\hspace{.1pt}\discretionary{.}{%
}{.}\hspace{.4pt}1016\discretionary{/}{%
}{/}0734\discretionary{%
}{-}{-}189X\discretionary{%
}{(}{(}83\discretionary{)}{%
}{)}90054\discretionary{%
}{-}{-}3}}


\bibitem{cooley1965algorithm}
J.~W. Cooley and J.~W. Tukey.
\newblock An algorithm for the machine calculation of complex fourier series.
\newblock {\em Mathematics of Computation}, 19(90):297--301, 1965. doi: {{%
10\hspace{.1pt}\discretionary{.}{%
}{.}\hspace{.4pt}1090\discretionary{/}{%
}{/}S0025\discretionary{%
}{-}{-}5718\discretionary{%
}{-}{-}1965\discretionary{%
}{-}{-}0178586\discretionary{%
}{-}{-}1}}


\bibitem{correll2012comparing}
M.~Correll, D.~Albers, S.~Franconeri, and M.~Gleicher.
\newblock Comparing averages in time series data.
\newblock In {\em ACM SIGCHI Conference on Human Factors in Computing Systems},
  pp. 1095--1104, 2012. doi: {{%
10\hspace{.1pt}\discretionary{.}{%
}{.}\hspace{.4pt}1145\discretionary{/}{%
}{/}2207676\hspace{.1pt}\discretionary{.}{%
}{.}\hspace{.4pt}2208556}}


\bibitem{correll2017regression}
M.~Correll and J.~Heer.
\newblock Regression by eye: Estimating trends in bivariate visualizations.
\newblock In {\em ACM SIGCHI Conference on Human Factors in Computing Systems},
  pp. 1387--1396, 2017. doi: {{%
10\hspace{.1pt}\discretionary{.}{%
}{.}\hspace{.4pt}1145\discretionary{/}{%
}{/}3025453\hspace{.1pt}\discretionary{.}{%
}{.}\hspace{.4pt}3025922}}


\bibitem{eeg_database}
A.~Delorme.
\newblock {EEG / ERP Public Database}.
\newblock
  \url{https://sccn.ucsd.edu/~arno/fam2data/publicly_available_EEG_data.html},
  Apr 2020.

\bibitem{douglas1973algorithms}
D.~H. Douglas and T.~K. Peucker.
\newblock Algorithms for the reduction of the number of points required to
  represent a digitized line or its caricature.
\newblock {\em Cartographica: The International Journal for Geographic
  Information and Geovisualization}, 10(2):112--122, 1973. doi: {{%
10\hspace{.1pt}\discretionary{.}{%
}{.}\hspace{.4pt}3138\discretionary{/}{%
}{/}FM57\discretionary{%
}{-}{-}6770\discretionary{%
}{-}{-}U75U\discretionary{%
}{-}{-}7727}}


\bibitem{2005eaton}
C.~Eaton, C.~Plaisant, and T.~Drizd.
\newblock Visualizing missing data: Graph interpretation user study.
\newblock In {\em IFIP Conference on Human-Computer Interaction}, pp. 861--872.
  Springer, 2005. doi: {{%
10\hspace{.1pt}\discretionary{.}{%
}{.}\hspace{.4pt}1007\discretionary{/}{%
}{/}11555261\_68}}


\bibitem{EdelsbrunnerHarer2010}
H.~Edelsbrunner and J.~Harer.
\newblock {\em Computational Topology: An Introduction}.
\newblock American Mathematical Society, Providence, RI, USA, 2010. doi: {{%
10\hspace{.1pt}\discretionary{.}{%
}{.}\hspace{.4pt}1090\discretionary{/}{%
}{/}mbk\discretionary{/}{%
}{/}069}}


\bibitem{gil1993computing}
J.~Gil and M.~Werman.
\newblock Computing 2-d min, median, and max filters.
\newblock {\em IEEE Transactions on Pattern Analysis and Machine Intelligence},
  15(5):504--507, 1993. doi: {{%
10\hspace{.1pt}\discretionary{.}{%
}{.}\hspace{.4pt}1109\discretionary{/}{%
}{/}34\hspace{.1pt}\discretionary{.}{%
}{.}\hspace{.4pt}211471}}


\bibitem{2019gogolou}
A.~{Gogolou}, T.~{Tsandilas}, T.~{Palpanas}, and A.~{Bezerianos}.
\newblock Comparing similarity perception in time series visualizations.
\newblock {\em IEEE Transactions on Visualization and Computer Graphics},
  25(1):523--533, 2019. doi: {{%
10\hspace{.1pt}\discretionary{.}{%
}{.}\hspace{.4pt}1109\discretionary{/}{%
}{/}TVCG\hspace{.1pt}\discretionary{.}{%
}{.}\hspace{.4pt}2018\hspace{.1pt}\discretionary{.}{%
}{.}\hspace{.4pt}2865077}}


\bibitem{hansen2011visualization}
C.~D. Hansen and C.~R. Johnson.
\newblock {\em Visualization handbook}.
\newblock Elsevier, 2011.

\bibitem{harrison2014ranking}
L.~Harrison, F.~Yang, S.~Franconeri, and R.~Chang.
\newblock Ranking visualizations of correlation using weber's law.
\newblock {\em IEEE Transactions on Visualization and Computer Graphics},
  20(12):1943--1952, 2014. doi: {{%
10\hspace{.1pt}\discretionary{.}{%
}{.}\hspace{.4pt}1109\discretionary{/}{%
}{/}TVCG\hspace{.1pt}\discretionary{.}{%
}{.}\hspace{.4pt}2014\hspace{.1pt}\discretionary{.}{%
}{.}\hspace{.4pt}2346979}}


\bibitem{2009heer}
J.~Heer, N.~Kong, and M.~Agrawala.
\newblock Sizing the horizon: The effects of chart size and layering on the
  graphical perception of time series visualizations.
\newblock In {\em ACM SIGCHI Conference on Human Factors in Computing Systems},
  pp. 1303--1312, 2009. doi: {{%
10\hspace{.1pt}\discretionary{.}{%
}{.}\hspace{.4pt}1145\discretionary{/}{%
}{/}1518701\hspace{.1pt}\discretionary{.}{%
}{.}\hspace{.4pt}1518897}}


\bibitem{holland1977robust}
P.~W. Holland and R.~E. Welsch.
\newblock Robust regression using iteratively reweighted least-squares.
\newblock {\em Communications in Statistics-theory and Methods}, 6(9):813--827,
  1977. doi: {{%
10\hspace{.1pt}\discretionary{.}{%
}{.}\hspace{.4pt}1080\discretionary{/}{%
}{/}03610927708827533}}


\bibitem{2010javed}
W.~{Javed}, B.~{McDonnel}, and N.~{Elmqvist}.
\newblock Graphical perception of multiple time series.
\newblock {\em IEEE Transactions on Visualization and Computer Graphics},
  16(6):927--934, 2010. doi: {{%
10\hspace{.1pt}\discretionary{.}{%
}{.}\hspace{.4pt}1109\discretionary{/}{%
}{/}TVCG\hspace{.1pt}\discretionary{.}{%
}{.}\hspace{.4pt}2010\hspace{.1pt}\discretionary{.}{%
}{.}\hspace{.4pt}162}}


\bibitem{kay2016beyond}
M.~Kay and J.~Heer.
\newblock Beyond weber's law: A second look at ranking visualizations of
  correlation.
\newblock {\em IEEE Transactions on Visualization and Computer Graphics},
  22(1):469--478, 2016. doi: {{%
10\hspace{.1pt}\discretionary{.}{%
}{.}\hspace{.4pt}1109\discretionary{/}{%
}{/}TVCG\hspace{.1pt}\discretionary{.}{%
}{.}\hspace{.4pt}2015\hspace{.1pt}\discretionary{.}{%
}{.}\hspace{.4pt}2467671}}


\bibitem{kerber2017geometry}
M.~Kerber, D.~Morozov, and A.~Nigmetov.
\newblock Geometry helps to compare persistence diagrams.
\newblock {\em Journal of Experimental Algorithmics (JEA)}, 22:1--4, 2017. doi:
  {{%
10\hspace{.1pt}\discretionary{.}{%
}{.}\hspace{.4pt}1145\discretionary{/}{%
}{/}3064175}}


\bibitem{gaussian}
S.~K. Kopparapu and M.~Satish.
\newblock Identifying optimal gaussian filter for gaussian noise removal.
\newblock In {\em Conference on Computer Vision, Pattern Recognition, Image
  Processing, and Graphics}, pp. 126--129, 2011. doi: {{%
10\hspace{.1pt}\discretionary{.}{%
}{.}\hspace{.4pt}1109\discretionary{/}{%
}{/}NCVPRIPG\hspace{.1pt}\discretionary{.}{%
}{.}\hspace{.4pt}2011\hspace{.1pt}\discretionary{.}{%
}{.}\hspace{.4pt}34}}


\bibitem{1989kosslyn}
S.~Kosslyn.
\newblock Understanding charts and graphs.
\newblock {\em Applied Cognitive Psychology}, 3:185 -- 225, 1989. doi: {{%
10\hspace{.1pt}\discretionary{.}{%
}{.}\hspace{.4pt}1002\discretionary{/}{%
}{/}acp\hspace{.1pt}\discretionary{.}{%
}{.}\hspace{.4pt}2350030302}}


\bibitem{noaa}
{National Centers for Environmental Information}.
\newblock {Climate Data Online: Web Services Documentation}.
\newblock \url{https://www.ncdc.noaa.gov/cdo-web/webservices/v2}, Apr 2020.

\bibitem{new_zealand}
{New Zealand Tourist Arrivals}.
\newblock \url{https://tradingeconomics.com/new-zealand/tourist-arrivals}, Apr
  2020.

\bibitem{nourbakhsh1994statistical}
M.~R. Nourbakhsh and K.~J. Ottenbacher.
\newblock The statistical analysis of single-subject data: a comparative
  examination.
\newblock {\em Physical Therapy}, 74(8):768--776, 1994. doi: {{%
10\hspace{.1pt}\discretionary{.}{%
}{.}\hspace{.4pt}1093\discretionary{/}{%
}{/}ptj\discretionary{/}{%
}{/}74\hspace{.1pt}\discretionary{.}{%
}{.}\hspace{.4pt}8\hspace{.1pt}\discretionary{.}{%
}{.}\hspace{.4pt}768}}


\bibitem{ondov2018face}
B.~Ondov, N.~Jardine, N.~Elmqvist, and S.~Franconeri.
\newblock Face to face: Evaluating visual comparison.
\newblock {\em IEEE Transactions on Visualization and Computer Graphics}, 2018.
  doi: {{%
10\hspace{.1pt}\discretionary{.}{%
}{.}\hspace{.4pt}1109\discretionary{/}{%
}{/}TVCG\hspace{.1pt}\discretionary{.}{%
}{.}\hspace{.4pt}2018\hspace{.1pt}\discretionary{.}{%
}{.}\hspace{.4pt}2864884}}


\bibitem{pandey2015deceptive}
A.~V. Pandey, K.~Rall, M.~L. Satterthwaite, O.~Nov, and E.~Bertini.
\newblock How deceptive are deceptive visualizations? an empirical analysis of
  common distortion techniques.
\newblock In {\em ACM SIGCHI Conference on Human Factors in Computing Systems},
  pp. 1469--1478, 2015. doi: {{%
10\hspace{.1pt}\discretionary{.}{%
}{.}\hspace{.4pt}1145\discretionary{/}{%
}{/}2702123\hspace{.1pt}\discretionary{.}{%
}{.}\hspace{.4pt}2702608}}


\bibitem{perreault2007median}
S.~Perreault and P.~H{\'e}bert.
\newblock Median filtering in constant time.
\newblock {\em IEEE Transactions on Image Processing}, 16(9):2389--2394, 2007.
  doi: {{%
10\hspace{.1pt}\discretionary{.}{%
}{.}\hspace{.4pt}1109\discretionary{/}{%
}{/}TIP\hspace{.1pt}\discretionary{.}{%
}{.}\hspace{.4pt}2007\hspace{.1pt}\discretionary{.}{%
}{.}\hspace{.4pt}902329}}


\bibitem{playfair1801commercial}
W.~Playfair.
\newblock {\em The commercial and political atlas: representing, by means of
  stained copper-plate charts, the progress of the commerce, revenues,
  expenditure and debts of england during the whole of the eighteenth century}.
\newblock T. Burton, 1801.

\bibitem{ramer1972iterative}
U.~Ramer.
\newblock An iterative procedure for the polygonal approximation of plane
  curves.
\newblock {\em Computer Graphics and Image Processing}, 1(3):244--256, 1972.
  doi: {{%
10\hspace{.1pt}\discretionary{.}{%
}{.}\hspace{.4pt}1016\discretionary{/}{%
}{/}S0146\discretionary{%
}{-}{-}664X\discretionary{%
}{(}{(}72\discretionary{)}{%
}{)}80017\discretionary{%
}{-}{-}0}}


\bibitem{rhodes1980generalized}
J.~D. Rhodes and S.~Alseyab.
\newblock The generalized chebyshev low-pass prototype filter.
\newblock {\em International Journal of Circuit Theory and Applications},
  8(2):113--125, 1980. doi: {{%
10\hspace{.1pt}\discretionary{.}{%
}{.}\hspace{.4pt}1002\discretionary{/}{%
}{/}cta\hspace{.1pt}\discretionary{.}{%
}{.}\hspace{.4pt}4490080205}}


\bibitem{suh2019topolines}
P.~Rosen, A.~Suh, C.~Salgado, and M.~Hajij.
\newblock {TopoLines: Topological Smoothing for Line Charts}.
\newblock In {\em EuroVis 2020 - Short Papers}, 2020. doi: {{%
10\hspace{.1pt}\discretionary{.}{%
}{.}\hspace{.4pt}2312\discretionary{/}{%
}{/}evs\hspace{.1pt}\discretionary{.}{%
}{.}\hspace{.4pt}20201053}}


\bibitem{2018entropy}
G.~Ryan, A.~Mosca, R.~Chang, and E.~Wu.
\newblock At a glance: Pixel approximate entropy as a measure of line chart
  complexity.
\newblock {\em IEEE Transactions on Visualization and Computer Graphics},
  25(1):872--881, 2019. doi: {{%
10\hspace{.1pt}\discretionary{.}{%
}{.}\hspace{.4pt}1109\discretionary{/}{%
}{/}TVCG\hspace{.1pt}\discretionary{.}{%
}{.}\hspace{.4pt}2018\hspace{.1pt}\discretionary{.}{%
}{.}\hspace{.4pt}2865264}}


\bibitem{2005horizon}
T.~Saito, H.~N. Miyamura, M.~Yamamoto, H.~Saito, Y.~Hoshiya, and T.~Kaseda.
\newblock Two-tone pseudo coloring: Compact visualization for one-dimensional
  data.
\newblock In {\em IEEE Symposium on Information Visualization (InfoVis)}, pp.
  173--180, 2005. doi: {{%
10\hspace{.1pt}\discretionary{.}{%
}{.}\hspace{.4pt}1109\discretionary{/}{%
}{/}INFVIS\hspace{.1pt}\discretionary{.}{%
}{.}\hspace{.4pt}2005\hspace{.1pt}\discretionary{.}{%
}{.}\hspace{.4pt}1532144}}


\bibitem{2018saket}
B.~{Saket}, A.~{Endert}, and A.~{Demiralp}.
\newblock Task-based effectiveness of basic visualizations.
\newblock {\em IEEE Transactions on Visualization and Computer Graphics},
  25(7):2505--2512, 2019. doi: {{%
10\hspace{.1pt}\discretionary{.}{%
}{.}\hspace{.4pt}1109\discretionary{/}{%
}{/}TVCG\hspace{.1pt}\discretionary{.}{%
}{.}\hspace{.4pt}2018\hspace{.1pt}\discretionary{.}{%
}{.}\hspace{.4pt}2829750}}


\bibitem{savitzky1964smoothing}
A.~Savitzky and M.~J. Golay.
\newblock Smoothing and differentiation of data by simplified least squares
  procedures.
\newblock {\em Analytical Chemistry}, 36(8):1627--1639, 1964. doi: {{%
10\hspace{.1pt}\discretionary{.}{%
}{.}\hspace{.4pt}1021\discretionary{/}{%
}{/}ac60214a047}}


\bibitem{shannon1949communication}
C.~E. Shannon.
\newblock Communication in the presence of noise.
\newblock {\em Proceedings of the Institute of Radio Engineers}, 37(1):10--21,
  1949. doi: {{%
10\hspace{.1pt}\discretionary{.}{%
}{.}\hspace{.4pt}1109\discretionary{/}{%
}{/}JRPROC\hspace{.1pt}\discretionary{.}{%
}{.}\hspace{.4pt}1949\hspace{.1pt}\discretionary{.}{%
}{.}\hspace{.4pt}232969}}


\bibitem{shao2016evaluation}
Y.~Shao, R.~S. Lunetta, B.~Wheeler, J.~S. Iiames, and J.~B. Campbell.
\newblock An evaluation of time-series smoothing algorithms for land-cover
  classifications using modis-ndvi multi-temporal data.
\newblock {\em Remote Sensing of Environment}, 174:258--265, 2016. doi: {{%
10\hspace{.1pt}\discretionary{.}{%
}{.}\hspace{.4pt}1016\discretionary{/}{%
}{/}j\hspace{.1pt}\discretionary{.}{%
}{.}\hspace{.4pt}rse\hspace{.1pt}\discretionary{.}{%
}{.}\hspace{.4pt}2015\hspace{.1pt}\discretionary{.}{%
}{.}\hspace{.4pt}12\hspace{.1pt}\discretionary{.}{%
}{.}\hspace{.4pt}023}}


\bibitem{2018szafir}
H.~Song and D.~Albers~Szafir.
\newblock Where's my data? evaluating visualizations with missing data.
\newblock {\em IEEE Transactions on Visualization and Computer Graphics},
  PP:1--1, 2018. doi: {{%
10\hspace{.1pt}\discretionary{.}{%
}{.}\hspace{.4pt}1109\discretionary{/}{%
}{/}TVCG\hspace{.1pt}\discretionary{.}{%
}{.}\hspace{.4pt}2018\hspace{.1pt}\discretionary{.}{%
}{.}\hspace{.4pt}2864914}}


\bibitem{szafir2018modeling}
D.~A. Szafir.
\newblock Modeling color difference for visualization design.
\newblock {\em IEEE Transactions on Visualization and Computer Graphics},
  24(1):392--401, 2018. doi: {{%
10\hspace{.1pt}\discretionary{.}{%
}{.}\hspace{.4pt}1109\discretionary{/}{%
}{/}TVCG\hspace{.1pt}\discretionary{.}{%
}{.}\hspace{.4pt}2017\hspace{.1pt}\discretionary{.}{%
}{.}\hspace{.4pt}2744359}}


\bibitem{2012templ}
M.~Templ, A.~Alfons, and P.~Filzmoser.
\newblock Exploring incomplete data using visualization techniques.
\newblock {\em Advances in Data Analysis and Classification}, 6:29--47, 2012.
  doi: {{%
10\hspace{.1pt}\discretionary{.}{%
}{.}\hspace{.4pt}1007\discretionary{/}{%
}{/}s11634\discretionary{%
}{-}{-}011\discretionary{%
}{-}{-}0102\discretionary{%
}{-}{-}y}}


\bibitem{1986tufte}
E.~R. Tufte.
\newblock {\em The Visual Display of Quantitative Information}.
\newblock Graphics Press, Cheshire, CT, USA, 1986. doi: {{%
10\hspace{.1pt}\discretionary{.}{%
}{.}\hspace{.4pt}1119\discretionary{/}{%
}{/}1\hspace{.1pt}\discretionary{.}{%
}{.}\hspace{.4pt}14057}}


\bibitem{us_bls}
{U.S. Bureau of Labor Statistics}.
\newblock \url{https://www.bls.gov/}, Apr 2020.

\bibitem{wang2018line}
Y.~Wang, F.~Han, L.~Zhu, O.~Deussen, and B.~Chen.
\newblock Line graph or scatter plot? automatic selection of methods for
  visualizing trends in time series.
\newblock {\em IEEE Transactions on Visualization and Computer Graphics},
  24(2):1141--1154, 2018. doi: {{%
10\hspace{.1pt}\discretionary{.}{%
}{.}\hspace{.4pt}1109\discretionary{/}{%
}{/}TVCG\hspace{.1pt}\discretionary{.}{%
}{.}\hspace{.4pt}2017\hspace{.1pt}\discretionary{.}{%
}{.}\hspace{.4pt}2653106}}


\bibitem{2018aspectratio}
Y.~{Wang}, Z.~{Wang}, L.~{Zhu}, J.~{Zhang}, C.~{Fu}, Z.~{Cheng}, C.~{Tu}, and
  B.~{Chen}.
\newblock Is there a robust technique for selecting aspect ratios in line
  charts?
\newblock {\em IEEE Transactions on Visualization and Computer Graphics},
  24(12):3096--3110, 2018. doi: {{%
10\hspace{.1pt}\discretionary{.}{%
}{.}\hspace{.4pt}1109\discretionary{/}{%
}{/}TVCG\hspace{.1pt}\discretionary{.}{%
}{.}\hspace{.4pt}2017\hspace{.1pt}\discretionary{.}{%
}{.}\hspace{.4pt}2787113}}


\bibitem{yahoo_finance}
{Yahoo Finance}.
\newblock \url{http://finance.yahoo.com/}, Apr 2020.

\bibitem{Zacks1999BarsAL}
J.~Zacks and B.~Tversky.
\newblock Bars and lines: A study of graphic communication.
\newblock {\em Memory \& cognition}, 27(6):1073--1079, 1999. doi: {{%
10\hspace{.1pt}\discretionary{.}{%
}{.}\hspace{.4pt}3758\discretionary{/}{%
}{/}BF03201236}}


\end{thebibliography}
